\documentstyle[amsfonts]{article}

\newcommand{\beq}{\begin{equation}}
\newcommand{\eeq}{\end{equation}}
\newcommand{\bea}{\begin{eqnarray}}
\newcommand{\eea}{\end{eqnarray}}

\newcommand{\Tint}{\int_{T^{2}}}
\newcommand{\ol}{\overline}

\newcommand{\cd}{\partial}

\newcommand{\p}{\wp}

\newcommand{\R}{{\Bbb R}}
\newcommand{\Z}{{\Bbb Z}}
\newcommand{\N}{{\Bbb N}}
\newcommand{\C}{{\Bbb C}}
\newcommand{\CP}{{\Bbb C}P^{1}}
\newcommand{\M}{M_{2}}
\newcommand{\wt}{\widetilde}
\newcommand{\wh}{\widehat}
\newcommand{\Mh}{\widehat{G}}
\newcommand{\Gh}{\widehat{G}}
\newcommand{\g}{\widetilde{g}}
\newcommand{\gt}{\widetilde{g}}
\newcommand{\gh}{\widehat{g}}
\newcommand{\ra}{\rightarrow}
\newcommand{\la}{\mbox{\boldmath $\lambda$}}
\newcommand{\ta}{\mbox{\boldmath $\tau$}}
\newcommand{\lll}{\scriptstyle \widehat{\lambda}}
\newcommand{\Rea}{{\rm Re}}

\newtheorem{theorem}{Theorem}
\newtheorem{lemma}{Lemma}

\begin{document}
\title{Lump dynamics in the $\CP$ model on the torus}
\author{J.M. Speight \\
Department of Mathematics \\
University of Texas at Austin \\
Austin, Texas 78712, U.S.A.}
\date{}
\maketitle

\begin{abstract}
The topology and geometry of the moduli space, $\M$, of degree 2 static 
solutions of the $\CP$ model on a torus (spacetime $T^{2}\times\R$) are
studied. It is proved that $\M$ is homeomorphic to the left coset space
$G/G_{0}$ where $G$ is a certain eight-dimensional noncompact Lie group and
$G_{0}$ is a discrete subgroup of order 4. Low energy two-lump dynamics is
approximated by geodesic motion on $\M$ with respect to a metric $g$ defined
by the restriction to $\M$ of the kinetic energy functional of the model.
This lump dynamics decouples into a trivial ``centre of mass'' motion and
nontrivial relative motion on a reduced moduli space. It is proved that
$(\M,g)$ is geodesically incomplete and has only finite diameter. A low
dimensional geodesic submanifold is identified and a full description of its
geodesics obtained.
\end{abstract}

\section{Introduction}
\label{sec:int}

The $\CP$ model in $(2+1)$ dimensions has long been popular in theoretical 
physics, both for its condensed matter applications, and as a simple nonlinear
field theory possessing topological solitons, usually called lumps. The
Euler-Lagrange equation of the system is not integrable, so there is no hope
of solving the multilump initial value problem exactly. Numerical simulations
of the model have revealed a rich diversity in the lump dynamics, which 
includes not only the now-familiar $90^{\circ}$ scattering in head on
collisions, but also lump expansion, collapse and singularity formation. It is
an interesting and highly nontrivial problem to understand the mechanisms
underlying this complicated dynamics.

Such understanding has been afforded in similar field theories (those of
Bogomol'nyi type) by the geodesic approximation of Manton 
\cite{Man1,AtiHit,Sam}. Here the low-energy dynamics of $n$ solitons is
approximated by geodesic motion in the moduli space of {\em static} $n$-soliton
solutions, $M_{n}$, the metric $g$ being defined by the restriction to $M_{n}$
of the kinetic energy functional of the field theory. So understanding
$n$-soliton dynamics is reduced to studying
 the topology and geometry of $(M_{n},g)$,
a finite dimensional, smooth Riemannian manifold.

Several authors have pursued this programme for the $\CP$ model in $\R^{2+1}$
with standard boundary conditions \cite{War,Lee}, concentrating on the case
of two lumps. There is, however, a technical problem: the metric on $\M$ does
not, strictly speaking, exist, that is, at every point $p\in\M$ some vectors in
$T_{p}\M$ are assigned
infinite length by the kinetic energy functional (they are ``non-normalizable
zero modes''). These divergences stem from the noncompactness of space 
$\R^{2}$. They are essentially due to the existence in the general static 
solution of scale and orientation parameters which are frozen in the geodesic
approximation because to alter them, no matter how slowly, costs infinite
kinetic energy. This is only possible because the kinetic energy is an
integral over a noncompact space. 
One can study geodesic motion orthogonal to the bad directions, or
one can remove the problem entirely by studying the model on a compact space
\cite{Spe}. In this paper we impose square periodic spatial boundary conditions
on the model, or, equivalently, place it on a flat torus. The aim 
is to establish rigorously results concerning the topology and geometry of
$(\M,g)$, and to describe their implications for low-energy two lump dynamics
on the torus, within the framework of the geodesic approximation. The work
is arranged as follows.

In section \ref{sec:setup} we introduce the $\CP$ model on the torus, and
review some relevant background material. In particular we use a standard
argument of Belavin and Polyakov to show that $\M$ is the space of degree
2 elliptic functions.

In section \ref{sec:mod} we equip $\M$ with a natural metric topology, and
prove that it is homeomorphic to the left coset space $G/G_{0}$, where
$G$ is the Lie group $PSL(2,\C)\times T^{2}$ and $G_{0}$ is discrete
subgroup of order 4. This allows one to give $\M$ a natural differentiable
structure (that of the smooth manifold $G/G_{0}$) and provides $\M$ with
a good global parametrization, using the covering space $G$.

In section \ref{sec:stat} this parametrization is used to survey the degree
2 static solutions and describe their energy density 
distributions. It is found that
exceptionally symmetric solutions exist with four, rather than two identical
energy lumps, as well as the expected two-lump and annular solutions.

In section \ref{sec:met} the metric $g$ on $\M$ is defined, and some of its
properties discussed. We lift $g$ to obtain $\g$, the metric on the covering 
space
$G$, and show that $\gt$ is a product metric on $PSL(2,\C)\times T^{2}$. In
this way, we show that lump dynamics in the geodesic approximation decomposes
into a trivial ``centre of mass'' motion, the $T^{2}$ part, and a
nontrivial relative motion, the $PSL(2,\C)$ part. So attention may be
restricted to geodesic motion on a reduced covering space, without loss of
generality.

In section \ref{sec:inc} it is proved that $(\M,g)$ is geodesically
incomplete by finding an explicit, maximally extended geodesic, and showing 
that it has only finite length. It follows that lumps can collapse to form
singularities in finite time.

In section \ref{sec:tgs} a 2-dimensional totally geodesic submanifold is
identified by computing the fixed point set of a discrete group of isometries.
The geodesics of this submanifold and their associated lump motions are 
described.

In section \ref{sec:dia} it is proved that $(\M,g)$ has only finite diameter,
despite its noncompactness. One should therefore visualize it as having only
finite extent. In consequence, all static solutions are close to the end of
moduli space, that is, close to collapse.

In section \ref{sec:conc} some concluding remarks are presented. Two 
3-dimensional totally geodesic submanifolds are identified, and it is shown
that $90^{\circ}$ head on scattering must occur in the model under certain
conditions. The present work is summarized, and extensions suggested.

\section{The $\CP$ model on the torus}
\label{sec:setup}

The field, a map from spacetime to $\CP$, $W:\R\times T^{2}\ra\CP$, will 
throughout be considered complex valued, so that we are using an inhomogeneous
coordinate on $\CP$, or equivalently, a stereographic coordinate on $S^{2}$,
exploiting the well known diffeomorphism between $\CP$ and the two sphere. The
metric and volume form on the codomain in terms of such a coordinate are,
respectively,
\beq
h=\frac{4\, du\, d\bar{u}}{(1+|u|^{2})^{2}}\qquad
\omega=\frac{2i\, du\wedge d\bar{u}}{(1+|u|^{2})^{2}}.
\eeq
It is convenient to use a complex coordinate on physical space also, by 
identifying $T^{2}$ with $\C/\Omega$ where $\Omega$ is the period module, 
which we choose, for concreteness, to be 
\beq
\Omega=\{n+im:n,m\in\Z\}.
\eeq
So we impose square periodic boundary conditions of unit period on $W$. 
Position in $T^{2}$ is parametrized by position $z=x+iy$ in the covering space
$\C$. The metric on spacetime is $\eta=dt^{2}-dx^{2}-dy^{2}$, and the action
functional of the field theory is the standard harmonic map functional
for mappings $(\R\times T^{2},\eta)\ra(\CP,h)$, that is,
\beq
S[W]=\int_{\R\times T^{2}}\frac{\cd_{\mu}W\cd_{\nu}\bar{W}}{(1+|W|^{2})^{2}}
\eta^{\mu\nu}.
\eeq
This may be written in a fashion reminiscent of Lagrangian mechanics,
$S=\int\, dt(T-V),$ upon definition of the kinetic and potential energy
functionals,
\bea
\label{kinetic}
T&=&\Tint\frac{|\dot{W}|^{2}}{(1+|W|^{2})^{2}} \\
V&=&\Tint\frac{1}{(1+|W|^{2})^{2}}\left(\left|\frac{\cd W}{\cd x}\right|^{2}+
\left|\frac{\cd W}{\cd y}\right|^{2}\right).
\eea

The configuration space $Q$ is $C^{1}(T^{2},S^{2})$, the space of continuously
differentiable maps $T^{2}\ra S^{2}$ (note that $V[W]$ is finite for all
$W\in Q$ by compactness of $T^{2}$). By Hopf's Degree Theorem \cite{Gui},
$Q$ decomposes into disjoint homotopy classes labelled by topological degree
$n$, an integer,
\beq
Q=\coprod_{n\in\Z}Q_{n}.
\eeq
Physically, $n$ is interpreted as the ``lump number'' of the configuration,
the excess of lumps over antilumps.

Static solutions are extremals of $V$, that is harmonic maps $T^{2}\ra S^{2}$.
The space of minimal energy static solutions in $Q_{n}$ is called the degree
$n$ moduli space, denoted $M_{n}$. A well-known argument due to Belavin and
Polyakov \cite{Bel} shows that $M_{n}$ ($n$ assumed nonnegative) is in fact
the space of degree $n$ elliptic functions, that is, holomorphic maps
$T^{2}\ra S^{2}$:
\bea
0&\leq&\Tint\frac{|\cd_{x}W+i\cd_{y}W|^{2}}{(1+|W|^{2})^{2}} \nonumber \\
&=& V[W]-\frac{1}{2}\Tint W^{*}\omega \nonumber \\
\label{6.5}
&=& V[W]-\frac{1}{2}{\rm Vol}(S^{2})n=V[W]-2\pi n
\eea
where $W^{*}\omega$ is the pullback of the volume from on $S^{2}$ by $W$.
It follows that
\beq
\label{7}
V|_{Q_{n}}\geq 2\pi n
\eeq
with equality if and only if $(\cd_{x}+i\cd_{y})W=0$, which is the 
Cauchy-Riemann equation for $W$. So if there exist degree $n$ elliptic
functions, then $M_{n}$ is the space of such functions, since any other 
function has higher energy. If there are no such functions, then $M_{n}$ is
empty, for the energy bound (\ref{7}) is optimal. To see this, consider the
following family of functions. For $\epsilon>0$ small, define $W_{\epsilon}\in
Q_{n}$ so that
\beq
W_{\epsilon}(z)=\left\{
\begin{array}{ll}
\frac{\epsilon^{2n}}{z^{n}} & |z|<\epsilon \\
 & \\
0 & |z|>2\epsilon \\
\end{array}\right.
\eeq
interpolating between these two regions with a smooth cutoff function. This
consists of a flat-space degree $n$ lump of width $\epsilon^{2}$ cut off on
a disc of radius $\epsilon$. Since $W_{\epsilon}$ is not exactly holomorphic,
$V[W_{\epsilon}]>2\pi n$, but the excess can be made arbitrarily small by
choosing $\epsilon$ small enough. 

It is easily proved that there are no unit degree elliptic functions 
\cite{Kno}, so we conclude that $M_{1}=\emptyset$, and the simplest nontrivial
moduli space is $M_{2}$.

\section{The degree two moduli space}
\label{sec:mod}

Weierstrass explicitly constructed a degree 2 elliptic function $\p$, and it
is on this that we base our parametrization of $\M$. The partial fraction
representation of $\p$ is
\beq
\label{9}
\p(z)=\frac{1}{z^{2}}-\sum_{\nu\in\Omega\backslash\{0\}}\left[
\frac{1}{(z-\nu)^{2}}-\frac{1}{\nu^{2}}\right].
\eeq
Several properties of $\p$ will be needed, some of which follow easily
from equation (\ref{9}), others of which are less straightforward. A 
comprehensive treatment can be found in \cite{Law}. Specifically:
\beq
\label{ident}
\begin{array}{ccc}
\p(iz)=-\p(z), & \p(-z)=\p(z), & \p(\bar{z})=\ol{\p(z)}, \\
 & & \\
 & \p'(z)^{2}=4\p(z)(\p(z)^{2}-e_{1}^{2}), & \\
\end{array}
\eeq
where $e_{1}=\p(\frac{1}{2})$ is a real number, approximately $6.875$. 
It follows that
$\p$ is real on
the boundary and central cross of the unit square, and purely imaginary on
the diagonals of the unit square (see figure 1) and that
 $\p$ has a double pole at
$0$ and a double zero at $(1+i)/2$.

Given one holomorphic function $\p\in\M$ one can obtain others by composing
on the right with a rigid translation of $T^{2}$ and on the left with a
M\"{o}bius transformation of $S^{2}$ since these preserve holomorphicity and
degree. In terms of a stereographic coordinate $W$ on $S^{2}$, M\"{o}bius 
transformations are unit degree rational maps \cite{Sch}
\beq
\label{10}
W\mapsto\frac{a_{11}W+a_{12}}{a_{21}W+a_{22}}
\eeq	
where $a_{ij}\in\C$ and $a_{11}a_{22}\neq a_{12}a_{21}$ else the degree
degenerates to zero. One may collect the parameters $a_{ij}$ into a matrix 
$M\in GL(2,\C)$ and denote the action of the matrix $M$ on $S^{2}$ defined
in equation (\ref{10}) by $W\mapsto M\odot W$. (The constraint 
$a_{11}a_{22}\neq a_{12}a_{21}$ is now $\det M\neq 0$, ensuring that $M$ is
invertible, and hence in $GL(2,\C)$.) Composition of M\"{o}bius transformations
coincides with matrix multiplication,
\beq
M_{2}\odot(M_{1}\odot W)\equiv (M_{2}M_{1})\odot W.
\eeq
Note, however, that this M\"{o}bius representation of $GL(2,\C)$ is not 
faithful since any pair of matrices $M,M'\in GL(2,\C)$ such that $M=\lambda
M'$ for some $\lambda\in\C$ generate the same M\"{o}bius transformation.
Denoting this scale equivalence $\sim$ we identify the M\"{o}bius group
with $GL(2,\C)/\sim$, each equivalence class of which may be represented by
a unimodular matrix ($\det M=1$). If $M$ is unimodular then so is $-M$, so
$SL(2,\C)$ is a double cover of $GL(2,\C)/\sim$, and the M\"{o}bius group
is identified with $SL(2,\C)/\Z_{2}$, usually denoted $PSL(2,\C)$, which is
easily seen to be six dimensional (the P stands for ``projective''). 

For the sake of brevity, let $G$ denote the eight dimensional Lie group
$PSL(2,\C)\times T^{2}$ with the group product $(M_{1},s_{1})\cdot(M_{2},s_{2})
=(M_{1}M_{2},s_{1}+s_{2})$.  We can define a $G$-action on $\M$, 
$G\times\M\ra\M$ such that $(g,W)\mapsto W_{g}$ where
\beq
W_{(M,s)}(z)=M\odot W(z-s).
\eeq
We claim that this action is transitive, since the $G$-orbit of $\p$ exhausts
$\M$.

\begin{lemma} For each $W\in\M$ there exists $(M,s)\in G$ such that
$W(z)=M\odot\p(z-s).$
\end{lemma}
{\bf Proof:} This may be 
established in several ways \cite{Kno2,DuV,Bea}. One economical, instructive
 (and
apparently novel) argument appeals to the Riemann-Hurwitz formula,
which constrains the number and valency of multivalent points of a holomorphic 
mapping between compact Riemann surfaces given their genera and the degree
of the map \cite{Bea2}. In the case of a degree 2 holomorphic map from $T^{2}$
to $S^{2}$ the formula states that any such function must have exactly 4
distinct double valency points (for $\p$ these are $0,\frac{1}{2},\frac{i}{2}$
and $(1+i)/2$). 

Let $W\in\M$ and $s\in T^{2}$ be one of its double valency points which is 
not a double pole. Then $(W(z+s)-W(s))^{-1}$ is another elliptic function with
a double pole at $0$, and no poles elsewhere (in the fundamental period
square). Its Laurent expansion about $0$ is
\beq
\frac{1}{W(z+s)-W(s)}=\frac{a_{1}}{z^{2}}+\frac{a_{2}}{z}+a_{3}+\cdots
\eeq
where $a_{1}\neq 0$. Consider $f(z)=[W(z+s)-W(s)]^{-1}-a_{1}\p(z)$. This is
an elliptic function with at most a simple pole at $0$, and no poles 
elsewhere. Hence it has degree $1$ or degree $0$. But there are no degree $1$
elliptic functions, and all degree $0$ elliptic functions are constant, so
$f(z)=c$. Defining
\beq
M'=\left(\begin{array}{cc}
a_{1}W(s) & cW(s)+1 \\
a_{1} & c \\
\end{array}\right)
\eeq
and $M=(\det M')^{-\frac{1}{2}}M'$, it follows that $W(z)=M\odot\p(z-s)$.
$\Box$

It is clear that for each $W\in\M$ the associated $g\in G$ is not unique,
since any one of the four distinct double valency points can be chosen as the
basis of the construction of $(M,s)$ outlined above. Conversely, given a choice
of $s\in T^{2}$, a double valency point of $W$, the construction of
$M\in PSL(2,\C)$ is unique, so for each $W\in\M$ there are exactly four
different $g\in G$ such that $W=\p_{g}$. In particular, we can construct
three alternative formulae for $\p(z)$ based on the three double valency
points $s_{0}=(1+i)/2$, $s_{1}=1/2$ and $s_{2}=i/2$ (the trivial formula
$\p(z)$ results from choosing $s=0$, the fourth double valency point):
\bea
\label{b}
\p(z)&\equiv&\frac{-e_{1}^{2}}{\p(z-s_{0})} \\
\label{bb}
&\equiv& \frac{e_{1}[\p(z-s_{1})+e_{1}]}{\p(z-s_{1})-e_{1}} \\
\label{bbb}
&\equiv&\frac{-e_{1}[\p(z-s_{2})-e_{1}]}{\p(z-s_{2})+e_{1}}.
\eea
These are found by computing the Laurent expansions of $\p$ about $s_{i}$
using (\ref{ident}), the formula for $\p'$. It is convenient to treat $\M$
as the $G$-orbit of $\p/e_{1}$, rather than $\p$. The identities
(\ref{b},\ref{bb},\ref{bbb}) can be rewritten
\beq
\frac{\p(z)}{e_{1}}\equiv U_{i}\odot\left(\frac{\p(z-s_{i})}{e_{1}}\right)
\qquad i=0,1,2
\eeq
where $U_{i}$ are the following $SU(2)$ matrices:
\beq
U_{0}=\left(\begin{array}{cc} 0 & 1 \\ -1 & 0 \\ \end{array}\right)\qquad
U_{1}=\frac{i}{\sqrt{2}}
\left(\begin{array}{cc} 1 & 1 \\  1 & -1 \\ \end{array}\right)\qquad
U_{2}=\frac{i}{\sqrt{2}}
\left(\begin{array}{cc} -1 & 1 \\ 1 & 1 \\ \end{array}\right).
\eeq
So the stabilizer of $\p/e_{1}$ under the $G$-action is
\beq
G_{0}=\{(I,0),(U_{0},s_{0}),(U_{1},s_{1}),(U_{2},s_{2})\},
\eeq
a discrete subgroup of $G$ isomorphic to the Viergruppe $V_{4}$, that is,
abelian, with each element its own inverse (when checking this recall that
$SL(2,\C)$ matrices which differ only in sign are identified). The $SU(2)$
subgroup of $SL(2,\C)$ acting on $S^{2}$ via $\odot$ is a double cover of
$SO(3)$ acting on $S^{2}$ via the natural rotation action. So $G_{0}$ is a
discrete group of simultaneous rotations of the target space $S^{2}$ and
translations of the domain $T^{2}$. In fact a straightforward calculation
shows that the $U_{i}$ are rotations of $S^{2}$ by $\pi$ about three
orthogonal axes.

For a general $W\in\M$, then, if $W=(\p/e_{1})_{g}$ then $W=(\p/e_{1})_{h}$
if and only if $h$ is an element of the left coset $gG_{0}$, which we will
henceforth denote $[g]$. So the mapping $\phi:G/G_{0}\ra\M$,
$\phi:[g]\mapsto\phi_{[g]}$ where
\beq
\phi_{[(M,s)]}(z)=M\odot\left(\frac{\p(z-s)}{e_{1}}\right)
\eeq
is well defined and bijective.  It would seem natural, therefore, to identify
$\M$ with $G/G_{0}$ via $\phi$, but this only makes sense provided $\phi$
is a homeomorphism. Before proving that this is indeed the case, there are
a few necessary preliminaries. Let $p:G\ra G/G_{0}$ be the projection map
$p(g)=[g]$. Since $G_{0}$ is a discrete subgroup of $G$, it acts freely
and properly discontinuously on $G$, so the quotient space $G/G_{0}$ is,
like $G$ itself, a Hausdorff, smooth manifold \cite{Thu}.
The pair $(G,p)$ is a covering
space of $G/G_{0}$, and $p$ is a local homeomorphism.

It is useful to define $\wt{\phi}:G\ra\M$ such that $\wt{\phi}=\phi\circ p$,
that is, $\wt{\phi}:g\mapsto\wt{\phi}_{g}=(\p/e_{1})_{g}$. The Lie group
$SL(2,\C)$ is noncompact, and is, in fact, homeomorphic to 
$\R^{3}\times SU(2)$, as may be shown \cite{Pen} by decomposing any 
$SL(2,\C)$ matrix $M$ into the product $HU$, where $U\in SU(2)$ and $H$ is a
positive definite, hermitian, unimodular matrix, this pair being unique.
The space of $H$-matrices is homeomorphic to $\R^{3}$ and may be parametrized
so that for all $\la\in\R^{3}$,
\beq
H(\la)=\sqrt{1+|\la|^{2}}I+\la\cdot\ta,
\eeq
where $\ta=(\tau_{1},\tau_{2},\tau_{3})$ are the Pauli spin matrices. It 
follows that $PSL(2,\C)\cong\R^{3}\times(SU(2)/\Z_{2})\cong\R^{3}\times SO(3)$,
and so $G\cong\R^{3}\times SO(3)\times T^{2}$.

To prove that $\phi$ is a homeomorphism we will need to understand the 
behaviour of $\wt{\phi}_{g}:T^{2}\ra S^{2}$ as $g$ approaches the end of $G$,
i.e.\ as $\lambda=|\la|\ra\infty$. For this purpose, consider the one
parameter family $\{\phi_{\lambda,\lll}=\wt{\phi}_{(M,0)}\in\M:M=H(\lambda
\widehat{\la}), \lambda\in(0,\infty)\}$ for some fixed $\widehat{\la}=\la/
\lambda\in S^{2}$. Explicitly,
\beq
\phi_{\lambda,\lll}(z)=H(\la)\odot\left(\frac{\p(z)}{e_{1}}\right).
\eeq
The action of $H(\la)$ on $S^{2}$ ($H(\la):W\mapsto H(\la)\odot W$) has 
exactly two
fixed points, $\widehat{\la}$ and $-\widehat{\la}$, and as $\lambda\ra\infty$
all but a vanishing neighbourhood of $-\widehat{\la}$ is mapped by $H(\la)$
to within a vanishing neighbourhood of $\widehat{\la}$ \cite{Spe}. So the
limiting function $\phi_{\infty,\lll}:T^{2}\ra S^{2}$ has the general form
\beq
\phi_{\infty,\lll}(z)=\lim_{\lambda\ra\infty}\phi_{\lambda,\lll}(z)=
\left\{\begin{array}{cc}
\widehat{\la} & z\notin(\p/e_{1})^{-1}(-\widehat{\la}) \\
-\widehat{\la} & z\in(\p/e_{1})^{-1}(-\widehat{\la}). \\
\end{array}\right. 
\eeq
That is, for generic $\widehat{\la}$, all but two points of $T^{2}$, the
preimages of $-\widehat{\la}$ under $\p/e_{1}$, are mapped by 
$\phi_{\infty,\lll}$ to $\widehat{\la}$, while these two points are mapped to
$-\widehat{\la}$. In the four special cases $\widehat{\la}\in\{(0,0,\pm 1),
(\pm 1,0,0)\}$, the preimages of $-\widehat{\la}$ coincide (double valency
points) so all but one point in $T^{2}$ is mapped to $\widehat{\la}$. 
The point to note is that in all cases $\phi_{\lambda,\lll}$ collapses to
a discontinuous limit.

The statement that $\phi:G/G_{0}\ra\M$ is a homeomorphism is, of course,
meaningless until we equip $\M$ with a topology (the domain inherits its
topology from $G$, which we take to have the natural product topology on
$PSL(2,\C)\times T^{2}$). There are many sensible choices for the topology on
$T^{2}$. One simple and directly physical choice is to endow $Q_{2}$ with the
metric topology where distance between configurations is measured by their
maximum pointwise deviation in the codomain $S^{2}$, so that $\M\subset Q_{2}$
inherits the relative topology. That is, let $d:S^{2}\times S^{2}\ra\R$ be the
usual distance function on $S^{2}$, and define $D:Q_{2}\times Q_{2}\ra\R$
such that, for all $W_{1},W_{2}\in Q_{2}$,
\beq
D(W_{1},W_{2})=\sup_{z\in T^{2}}d(W_{1}(z),W_{2}(z)).
\eeq
It is straightforward to verify that $D$ satisfies the axioms of a distance
function. The resulting metric topology on $Q_{2}$ is Hausdorff, as is any
metric topology \cite{Cho1}. Rather than break up the smooth manifold $G$
into coordinate charts, it is convenient to equip $G$ with a metric topology
also, as follows: let $h$ be the (Riemannian) product metric
\beq
h=(d\la\cdot d\la)\oplus h_{SO(3)}\oplus ds\, d\bar{s}
\eeq
on $G\cong \R^{3}\times SO(3)\times T^{2}$, where $h_{SO(3)}$ is the 
biinvariant metric on $SO(3)$ of unit volume. The Riemannian manifold $(G,h)$
has a natural distance function $\wt{d}$ where $\wt{d}(g_{1},g_{2})$ is the
infimum of lengths (with respect to $h$) of piecewise $C^{1}$ paths connecting 
$g_{1}$ and $g_{2}$. That $\wt{d}$ is a distance function, and that the
associated metric topology coincides with the original topology on $G$
(independent of the choice of $h$) are standard theorems of Riemannian 
geometry \cite{Gal}. We may now state and prove Theorem 1. Throughout,
$B_{\epsilon}(x)$ denotes the open ball of radius $\epsilon$ centred on
$x$, where the space containing $x$ ($S^{2}$, $\M$ or $G$), and hence the 
appropriate distance
function ($d$, $D$ or $\wt{d}$), should be clear from context.

\begin{theorem} The bijection $\phi:G/G_{0}\ra\M$ is a homeomorphism.
\end{theorem}
{\bf Proof:} We must prove that both $\phi$ and $\phi^{-1}$ are continuous.
To prove the former, it suffices to show that $\wt{\phi}=\phi\circ p$, 
 is continuous, since the projection $p$ is a local
homeomorphism. Fix $g_{0}\in G$ and $\epsilon>0$. Then we must show that
$\exists \delta>0$ such that $\forall g\in B_{\delta}(g_{0})$, 
$\wt{\phi}_{g}\in
B_{\epsilon}(\wt{\phi}_{g_{0}}).$
Let $\phi_{*}:G\times T^{2}\ra S^{2}$ such that 
$\phi_{*}(g,z)=\wt{\phi}_{g}(z).$ Note that $\phi_{*}$ is manifestly 
continuous. Hence, for each $\tilde{z}\in T^{2}$ there exists 
$\delta(\tilde{z})>0$
such that
\beq
\label{a}
(g,z)\in B_{\delta(\tilde{z})}(g_{0})\times B_{\delta(\tilde{z})}(\tilde{z})
\Rightarrow
d(\phi_{*}(g,z),\phi_{*}(g_{0},\tilde{z}))<\frac{\epsilon}{3}.
\eeq
The collection of open balls $\{B_{\delta(z)}\subset T^{2}:z\in T^{2}\}$ is an
open cover of $T^{2}$. Since $T^{2}$ is compact, there exists a finite 
subcover $\{B_{\delta(z_{j})}(z_{j}):j=1,2,\ldots,N\}$. Define $\delta=
\inf\{\delta(z_{j}):j=1,2,\ldots,N\}>0$.

Now, let $g\in B_{\delta}(g_{0})$ and consider 
$D(\wt{\phi}_{g},\wt{\phi}_{g_{0}})$. For each $z\in T^{2}$ there exists
$j\in\{1,2,\ldots,N\}$ such that $z\in B_{\delta(z_{j})}(z_{j})$. Further,
$g,g_{0}\in B_{\delta}(g_{0})\subset B_{\delta(z_{j})}(g_{0})$ by definition of
$\delta$, so $(g,z),(g_{0},z)\in B_{\delta(z_{j})}(g_{0})\times
B_{\delta(z_{j})}(z_{j})$. Hence, using (\ref{a}) and the triangle inequality,
\beq
d(\phi_{*}(g,z),\phi_{*}(g_{0},z))\leq
d(\phi_{*}(g,z),\phi_{*}(g_{0},z_{j}))+
d(\phi_{*}(g_{0},z),\phi_{*}(g_{0},z_{j}))<\frac{2\epsilon}{3}.
\eeq
Thus,
\beq
D(\wt{\phi}_{g},\wt{\phi}_{g_{0}})=\sup_{z\in T^{2}}
d(\phi_{*}(g,z),\phi_{*}(g_{0},z))\leq\frac{2\epsilon}{3}<\epsilon,
\eeq
so $\wt{\phi}$ is continuous.

To prove that $\phi^{-1}$ is continuous we again convert the problem to one
involving $\wt{\phi}$, using general properties of covering spaces. Fix
$[g_{0}]\in G/G_{0}$ and choose any open neighbourhood $U$ of $[g_{0}]$.
The inverse image of $[g_{0}]$ under $p$ is the left coset $g_{0}G_{0}=
\{g_{0},g_{1},g_{2},g_{3}\}$. Since $p$ is a local homeomorphism there exists
$\epsilon>0$ such that $p(U_{\epsilon})\subset U$, where 
\beq
U_{\epsilon}:=\bigcup_{i=0}^{3}B_{\epsilon}(g_{i})\subset G.
\eeq
We will show that there exists $\delta>0$ such that 
$\wt{\phi}^{-1}(B_{\delta}(W_{0}))\subset U_{\epsilon}$, where $W_{0}=
\phi_{[g_{0}]}\in\M$. It follows that $\phi^{-1}(B_{\delta}(W_{0}))\subset U$,
and hence that $\phi^{-1}$ is continuous. 

For each $n\in\N$, define the compact set
\beq
A_{n}=\ol{B}_{n}(I,0)\backslash U_{\epsilon}\subset G
\eeq
where $\ol{B}_{n}(I,0)$ is the closed ball of radius $n$ centred on $(I,0)\in
G$. Since $\wt{\phi}$ is continuous, $\wt{\phi}(A_{n})\subset\M$ is also
compact, and therefore closed ($\M$ is Hausdorff). Hence the complement 
$\M\backslash\wt{\phi}(A_{n})$ is open, and it contains $W_{0}$ by construction
(since $A_{n}\cap U_{\epsilon}=\emptyset$), so there exists $\delta_{n}>0$ such
that $B_{\delta_{n}}(W_{0})\subset\M\backslash\wt{\phi}(A_{n})$. In this way,
construct a positive sequence $(\delta_{n})_{n=1}^{\infty}$, which, without
loss of generality, we may assume is decreasing and converges to $0$. Consider
the preimage of $B_{\delta_{n}}(W_{0})$ under $\wt{\phi}$. By construction,
$\wt{\phi}^{-1}(B_{\delta_{n}}(W_{0}))\cap A_{n}=\emptyset$, so every point in
the preimage lies either in $U_{\epsilon}$, or at a distance greater than
$n$ from $(I,0)\in G$.

We claim that there exists $N\in\N$ such that
$\wt{\phi}^{-1}(B_{\delta_{n}}(W_{0}))\subset U_{\epsilon}$. Choosing $\delta=
\delta_{N}$, the proof is then complete. Assume this claim is false. Then
$\forall n\in\N$ there exists $g_{n}\notin\ol{B}_{n}(I,0)$ such that
$\wt{\phi}_{g_{n}}\in B_{\delta_{n}}(W_{0})$. For each $n$, choose such a 
$g_{n}$ and consider the sequence $(g_{n})_{n=1}^{\infty}$. Since
$(\delta_{n})_{n=1}^{\infty}\ra 0$, the image of the sequence under 
$\wt{\phi}$, $(\wt{\phi}_{g_{n}})_{n=1}^{\infty}$ converges to $W_{0}$ in $\M$.
Define two projection maps on $G\cong\R^{3}\times SO(3)\times T^{2}$:
\bea
\pi_{1}:G\ra[0,\infty)\qquad\mbox{such that} & & \pi_{1}(\la,U,s)=\lambda=|\la|
\nonumber \\
\pi_{2}:G\ra S^{2}\times SO(3)\times T^{2}\qquad\mbox{such that} & &
\pi_{2}(\la,U,s)=(\widehat{\la},U,s).
\eea
The singularity of $\pi_{2}$ when $\lambda=0$ is irrelevant here.  By 
construction, $(\lambda_{n})_{n=1}^{\infty}=(\pi_{1}(g_{n}))_{n=1}^{\infty}$
is unbounded, and without loss of generality, we may choose $g_{n}$ such that
$\lambda_{n}$ is increasing.  Since
$(\pi_{2}(g_{n}))_{n=1}^{\infty}$ takes values in a compact space, it has,
by the Bolzano-Weierstrass theorem, a convergent subsequence
$(\pi_{2}(g_{n_{r}}))_{r=1}^{\infty}$. By translation and rotation symmetry
of $T^{2}$ and $S^{2}$ respectively, we may assume without loss of
generality that its limit is $(\widehat{\la},I,0)$. 

Consider the image under $\wt{\phi}$ of the associated subsequence
$(g_{n_{r}})_{r=1}^{\infty}$, which approaches the end of $\R^{3}\times
SO(3)\times T^{2}$ asymptotic to the line 
$\{(t\widehat{\la},I,0):t\in(0,\infty)\}$. The function 
$\wt{\phi}_{g_{n_{r}}}(z)$
converges pointwise, as $r\ra\infty$, to 
$\phi_{\infty,{\scriptstyle\widehat{\lambda}}}(z)$, the
limiting function previously described (to check this, use continuity of
the $SO(3)$ and $T^{2}$ actions on $S^{2}$ and $T^{2}$ respectively,
and of the function $\p/e_{1}$). But 
$\phi_{\infty,{\scriptstyle \widehat{\lambda}}}$, being
discontinuous, cannot be in $\M$, and hence cannot be $W_{0}$, a contradiction.
$\Box$

\section{Degree 2 static solutions}
\label{sec:stat}

An immediate corollary of Theorem 1 is that $(G,\wt{\phi})$ is a covering
space of $\M$. The aim of this section is to describe the connexion between
any point $g\in G$ and its corresponding static solution $\wt{\phi}_{g}\in\M$,
that is, to obtain a picture of what the static lumps look like, and how
they change as $g$ varies. A configuration $W$ may be visualized as a 
distribution of unit length three-vectors (``arrows'') over the torus. The
energy density function of $W$ is
\beq
{\cal E}(x,y)=\frac{|W_{x}|^{2}+|W_{y}|^{2}}{(1+|W|^{2})^{2}},
\eeq
so the energy is located where the direction of the arrows is varying
sharply in $(x,y)$, in other words, where neighbouring arrows are stretched
apart. It is the function ${\cal E}$ that we will describe as $W$ varies in
$\M$.

For this purpose, rather than using the hermitian-unitary (or ``polar'')
decomposition of $SL(2,\C)$ used above, another standard decomposition is
convenient. Namely, any $M\in SL(2,\C)$ may be uniquely decomposed into a
product $UT$ with $U\in SU(2)$ and $T$ upper triangular, real on the diagonal,
positive definite and unimodular. The space of such $T$-matrices is 
homeomorphic to $\R^{+}\times\C$ (here $\R^{+}=(0,\infty)$) and may be
parametrized thus:
\beq
T(\alpha,\rho)=\left(\begin{array}{cc}
\sqrt{\alpha e_{1}} & \sqrt{\alpha e_{1}}\rho \\
0 & 1/\sqrt{\alpha e_{1}} \\
\end{array}\right).
\eeq
This allows one to write any $W\in\M$ in the form
\beq
W(z)=(UT)\odot\left(\frac{\p(z-s)}{e_{1}}\right)=U\odot[\alpha(\p(z-s)+\rho)].
\eeq
Changing $U\in SU(2)$ merely produces a global internal rotation of the
solution and so has no effect on ${\cal E}(z)$. Changing $s\in T^{2}$ 
translates the solution on the torus, so it suffices to examine the three
parameter family
\beq
W(z)=\alpha(\p(z)+\rho)
\eeq
$(\alpha,\rho)\in\R^{+}\times\C$, whose energy density is
\beq
\label{energy}
{\cal E}(z)=\frac{8\alpha^{2}|\p(z)||\p(z)^{2}-e_{1}^{2}|}{(1+\alpha^{2}|\p(z)
+\rho|^{2})^{2}}.
\eeq
Note that for all $(\alpha,\rho)$, ${\cal E}=0$ at the four double valency
points $z=0,s_{0},s_{1},s_{2}$, around which the direction of the arrows is
constant to first order. 

The behaviour of ${\cal E}$ as $(\alpha,\rho)$ covers $\R^{+}\times\C$ is
remarkably varied, going beyond the two-lump and annular structures one might
expect by analogy with the planar $\CP$ model. First, consider the case
$\alpha=1$. Here, the energy is located in lumps close to the two roots of
$\p(z)+\rho$ (symmetrically placed about $s_{0}$ since $\p$ is even) where the
denominator of (\ref{energy}) is smallest. The only exceptions are when these 
roots coincide, $\rho=0,-e_{1},e_{1}$, or are close to coincidence, for then
the lumps lose their individual identity and form a ring-like structure
(centred on $s_{0}$, $s_{1}$ or $s_{2}$ respectively) rather reminiscent of
coincident planar solitons (figure 2). 
If we now imagine increasing $\alpha$ above $1$,
the effect on $W$ is to pull all the arrows in the configuration towards the
north pole of $S^{2}$ ($W=\infty$), so that those close to the south pole
($W=0$) are stretched apart. Since the energy is located where the arrows
are stretched apart, increasing $\alpha$ therefore tends to concentrate ${\cal
E}$ more strongly on roots of $\p(z)+\rho$, and the lumps become taller and
narrower. As $\alpha\ra\infty$ the lumps collapse and ``pinch off''.
Conversely, if $\alpha$ is decreased below $1$ the arrows of the configuration
are pulled southwards, and for $\alpha$ very small, ${\cal E}$ concentrates
on the double pole of $\p(z)+\rho$ ($z=0$), where $W$ points north. In this
case a ring structure appears, centred on $z=0$, and collapses to zero width
as $\alpha\ra 0$. These two cases are compared in figure 3.
Noting the symmetry property $\p(iz)\equiv-\p(z)$, we see 
that whenever $\rho$ passes through $0\in\C$ along a smooth curve, the
roots of $\p(z)+\rho$ coalesce and emerge at right angles to their line
of approach, giving a first hint that the familiar $90^{\circ}$ scattering
of lumps through a ring structure may take place in the geodesic approximation.
We shall return to this point later.

The special case $\rho=0$ is exceptional, and will be prominent in later
sections. Examining the formula (\ref{energy}) in this case, we see that the
global maxima of ${\cal E}$ must occur where $\p$ is purely imaginary. If not,
assume that a global maximum occurs at $z_{0}$ and let $\p(z_{0})=u\in\C
\backslash i\R$. Then 
\beq
{\cal E}(z_{0})=
\frac{8\alpha^{2}|u||u^{2}-e_{1}^{2}|}{(1+\alpha^{2}|u|^{2})^{2}}<
\frac{8\alpha^{2}|u|(|u|^{2}+e_{1}^{2})}{(1+\alpha^{2}|u|^{2})^{2}}
\eeq
where the inequality is strict since $u^{2}$ is not real-negative. But
there exists $z_{1}\in T^{2}$ such that $\p(z_{1})=i|u|$, and
\beq
{\cal E}(z_{1})=
\frac{8\alpha^{2}|u||-|u|^{2}-e_{1}^{2}|}{(1+\alpha^{2}|u|^{2})^{2}}>
{\cal E}(z_{0}),
\eeq
a contradiction. Given the symmetry of ${\cal E}$ under $\p\mapsto -\p$, and
that $\p$ is even, it follows that ${\cal E}(z)$ has at least four peaks on the
diagonals of the unit square, symmetrically placed about $s_{0}$. Plots of
${\cal E}$ confirm that there are, in fact, exactly four such peaks
(figure 4). The
must symmetric case is $\alpha=1/e_{1}$, that is, $W(z)=\p(z)/e_{1}$. Here,
using the identity (\ref{b}), one can easily show that ${\cal E}(z-s_{0})\equiv
{\cal E}(z)$, so the four peaks are located halfway towards the centre
$s_{0}$ along the diagonals, i.e.\ at the points $(1+i)/4,(3+i)/4,3(1+i)/4,
(1+3i)/4$. This solution has the most evenly spread energy distribution
possible. Once again, one can consider the effect of increasing $\alpha$
(pulling the arrows northwards) or decreasing $\alpha$ (pulling southwards)
for this family. Increasing $\alpha$ moves the lumps towards $s_{0}$ where they
coalesce, form a shrinking ring structure and pinch off. Decreasing $\alpha$
has the same effect, except the ring is centred on $0$ rather than $s_{0}$.
In fact, the solution $\alpha\p(z)$ is identical, up to the rotation and 
translation $(U_{0},s_{0})\in G_{0}$, to the solution 
$\p(z)/(e_{1}^{2}\alpha)$.

When $\alpha$ is close to $1/e_{1}$
and $|\rho|$ is small but nonzero, the behaviour of ${\cal E}(z)$ is 
intermediate between the two cases described above. It has four peaks, but
two of these are larger than the other two (figure 5).

\section{The metric on $\M$}
\label{sec:met}

The argument of Belavin and Polyakov (\ref{6.5}) shows that $\M$ is the flat
valley bottom of $Q_{2}$, on which $V$ attains its topological minimum value,
$4\pi$. Any departure from $\M$ involves increasing $V$, and hence climbing
the valley walls. Consider the initial value problem where $W$ starts on $\M$
and is given a small push tangential to it. Then, by energy conservation, it
must stay close to $\M$ during its subsequent evolution. In the geodesic
approximation one constrains the configuration to lie on $\M$ for all time,
but allows the position in $\M$ to evolve in time according to the constrained
action principle. Since $V=4\pi$ always, the dynamics is determined solely by
the kinetic energy functional (\ref{kinetic}). Using the homeomorphism
$\phi$ we can transfer the differentiable structure of $G/G_{0}$ to $\M$.
Let $\{q_{i}:i=1,2,\ldots,8\}$ be local coordinates on $\M$, and consider the
kinetic energy  $T$ as $q^{i}$ vary in time:
\beq
T=g_{ij}(q)\dot{q}^{i}\dot{q}^{j}
\eeq
where
\beq
\label{20}
g_{ij}(q)=\Rea\Tint\frac{1}{(1+|W|^{2})^{2}}\frac{\cd W}{\cd q_{i}}
\frac{\cd \ol{W}}{\cd q^{j}}.
\eeq
Equation (\ref{20}) defines a Riemannian metric on $\M$,
$g=g_{ij}dq^{i}dq^{j}$, and furthermore the constrained Euler-Lagrange
equation (obtained by varying the action $S[q]=\int dt\, T(q,\dot{q})$) is
the geodesic equation for $(\M,g)$. The conjecture is, then, that geodesics in
this Riemannian manifold are, when travelled at low speed, close to low-energy
two-lump dynamical
solutions of the $\CP$ model. Some justification for this can be
found when comparison is made with other models for which the approximation
has been used. In the case of abelian-Higgs vortices, for example,
the approximation \cite{Sam} is supported by rigorous analysis \cite{Stu} and 
extensive
numerical solution of the full field equations \cite{Mye}.

Ideally, one would like an explicit, closed-form expression for the metric
$g$, but this is rarely possible in practice. There are exceptions
\cite{AtiHit,Spe,Str}, but unfortunately this is not one of them. It is
possible to place fairly strong constraints on the possible form of $g$, but
not to write it down explicitly (to do so requires, naively, the evaluation
of 36 integrals over $T^{2}$, each with 8 parameters). It is convenient to
lift the geometry to the covering space $(G,\gt)$ where $\gt=\wt{\phi}^{*}g$,
the pullback of the metric $g$ by the covering projection $\wt{\phi}$.
The most useful constraint is that $\gt$ is a product metric 
$\gt=\gh\oplus\delta$ on $PSL(2,\C)\times T^{2}$ where $\delta=2\pi ds\, 
d\bar{s}$. By product metric \cite{Gal2}
we mean block diagonal with $\gh$ independent of
position in $T^{2}$ and $\delta$ independent of position in $PSL(2,\C)$.

This is easily established if we recall that any $W\in\M$ is a rational
function of $\p(z-s)$, so denoting by $\mu$ any one of the six $PSL(2,\C)$
moduli, we see that
\bea
\frac{\cd W}{\cd s}&=&-R_{1}(\p(z-s))\p'(z-s) \nonumber \\
\frac{\cd W}{\cd\mu}&=&R_{2}(\p(z-s))
\eea
where $R_{1}$, $R_{2}$ are rational functions. So the $(\mu,s)$ component of
$\gt$ is
\beq
\gt_{\mu s}=\Rea\Tint f(\p(z-s))\p'(z-s)=\Rea\Tint f(\p(z))\p'(z)=0
\eeq
since $\p$ is even while $\p'$ is odd (here $f(u)=-(1+|R_{1}(u)|^{2})^{-2}
\ol{R_{1}(u)}R_{2}(u)$). Similarly, $\gt_{\mu\bar{s}}=0$, so $\gt$
is block diagonal as claimed. Translation symmetry implies that $\gt$,
and hence $\gh$, must be independent of $s$. Hence, it remains to show that
$\delta$ is independent of the $PSL(2,\C)$ moduli. Let $W(x,y,t)=
\wt{W}(z-s(t))$, $\wt{W}\in\M$, and compute the kinetic energy,
\beq
T=\Tint\frac{|\wt{W}'|^{2}}{(1+|\wt{W}|^{2})^{2}}|\dot{s}|^{2}=
\frac{1}{2}V[\wt{W}]|\dot{s}|^{2}=2\pi|\dot{s}|^{2}.
\eeq
Since this is, by definition, $g_{s\bar{s}}|\dot{s}|^{2}$, we read off the
metric $\delta=2\pi ds\, d\bar{s}$ on $T^{2}$. 

The geodesic equation for $(G,\gt)$ decouples into independent
geodesic equations for $(PSL(2,\C),\gh)$ and $(T^{2},\delta)$. Consequently,
we may identify $s$ as an effective ``centre of mass coordinate'' which
drifts on $T^{2}$ at constant velocity, independent of the lumps'
relative motion in $PSL(2,\C)$. Without loss of generality, therefore,
we can investigate geodesic motion in the reduced covering space $(\Gh,\gh)$,
where $\Gh$ denotes $PSL(2,\C)$. So lump dynamics in the geodesic approximation
has Galilean boost symmetry. This may be understood as a remnant of the
Lorentz symmetry of the $\CP$ model in $\R^{2+1}$: the field equation is still
Lorentz invariant, but the spatial boundary conditions now are not. Under
a Lorentz boost, they suffer Lorentz contraction. In a low speed approximation
such as this, however, Galilean boost symmetry is recovered, since the spatial
contraction is a high order effect.

One further constraint on $\gh$ will prove
useful: since the kinetic energy functional is invariant under global internal
rotations of $W$ (rotations of the codomain $S^{2}$), $SU(2)$ \
acts isometrically
by left multiplication on $(\Gh,\gh)$ \cite{Spe}. Briefly, $\gh$ (or $\gt$)
is {\em left-invariant} under $SU(2)$. 

\section{Geodesic incompleteness of $(\M,g)$}
\label{sec:inc}

One of the most basic questions one can ask about a Riemannian manifold 
without boundary is whether it is geodesically complete, that is, whether
all geodesics can be extended infinitely in time (forwards and backwards).
In view of the noncompactness of $\M$, this is a nontrivial question for
$(\M,g)$. We will prove that $(\M,g)$ is , in fact, geodesically incomplete,
by finding a geodesic which, although maximally extended, has only finite
length (since geodesics are traversed at constant speed, this is sufficient).
This geodesic is obtained explicitly, despite our lack of explicit
information about $g$, by using discrete isometries to identify a one
dimensional geodesic submanifold. Such arguments have been used to obtain
multimonopole scattering geodesics
\cite{Man2} given similarly scant knowledge of the metric on moduli space.

The key observation is that the fixed point set of a discrete group of
isometries of a Riemannian manifold is (if a submanifold) a totally geodesic
submanifold, that is, a geodesic which starts on and tangential to the
fixed point set must remain on the fixed point set for all subsequent time.
This follows directly from uniqueness of solutions to the initial value
problem of an ordinary differential equation. 
If a discrete group is found whose fixed point set is diffeomorphic to $\R$, 
then the set itself is a geodesic.

The following mappings are isometries of $(\Gh,\gh)$,
\bea
P:M&\mapsto&\ol{M} \\
R:M&\mapsto&\tau_{3}M\tau_{3}.
\eea
To see this, consider their effect on $W(z)=M\odot(\p(z)/e_{1})$,
\bea
P:W(z)&\mapsto&\ol{M}\odot\frac{\p(z)}{e_{1}}=\ol{M}\odot
\frac{\ol{\p(\bar{z})}}{e_{1}}=\ol{W(\bar{z})} \\
R:W(z)&\mapsto&(\tau_{3}M\tau_{3})\odot\frac{\p(z)}{e_{1}}
=-[M\odot(-\p(z)/e_{1})]
=-W(iz).
\eea
So $P$ produces simultaneous reflexions in both domain ($z\mapsto\bar{z}$) and
codomain ($W\mapsto\ol{W}$), while $R$ produces rotations of $\pi/2$
in the domain ($z\mapsto iz$) and $\pi$ in the codomain ($W\mapsto-W$), all of
which are symmetries of the $\CP$ model. The composition of $P$ and $R$ in
either order (they commute) is another isometry of $(\Gh,\gh)$, Since
$P^{2}=R^{2}=(PR)^{2}=Id$, the isometries $\{Id,P,R,PR\}$ form the Viergruppe
$V_{4}$ under composition. A straightforward calculation shows that
$\wh{\Sigma}_{V}$, the fixed point set of $V_{4}$ is
\beq
\wh{\Sigma}_{V}=\{{\rm diag}((\alpha e_{1})^{\frac{1}{2}},
(\alpha e_{1})^{-\frac{1}{2}}):\alpha\in\R^{+}\}.
\eeq
This is clearly a submanifold of $\Gh$ diffeomorphic to $\R$, and hence is a
geodesic. Its image under the projection $\wt{\phi}$ is
$\Sigma_{V}=\{\alpha\p(z):\alpha\in\R^{+}\}$ which 
is a geodesic of $(\M,g)$, also
diffeomorphic to $\R$. 

The submanifold $\Sigma_{V}$ was described at the end of section 
\ref{sec:stat}. The lump motion corresponding to this geodesic is an
infinitely tall thin ring centred at $0$ in the past spreading out into
four distinct identical peaks, which recombine to form an infinitely tall thin
ring centred on $s_{0}$ in the future (assuming that $\Sigma_{V}$ is traversed
in the sense of increasing $\alpha$). The question remains whether $\Sigma_{V}$
is traversed in finite time, i.e.\ has finite length, and to answer this one
needs to understand the induced metric $g_{V}$ on $\Sigma_{V}$. The restriction
of $g$ to $\Sigma_{V}$ is
\beq
g_{V}=f(\alpha)d\alpha^{2}
\eeq
where
\beq
\label{fdef}
f(\alpha)=\Tint\frac{|\p|^{2}}{(1+\alpha^{2}|\p|^{2})^{2}}.
\eeq
Note that $f$ is clearly positive and decreasing, and is easily shown to have
limits $\infty$ and $0$ as $\alpha\ra 0$ and $\alpha\ra\infty$
respectively.
To prove that $\Sigma_{V}$ has finite length we will need detailed asymptotic
estimates for $f$ in these two limits.  The identity (\ref{b}) implies that
\beq
\label{fident}
f(\alpha)\equiv\frac{1}{(\alpha e_{1})^{4}}f(\frac{1}{\alpha e_{1}^{2}}),
\eeq
so the behaviour in one limit follows directly from the behaviour in the
other.

\begin{lemma} The following asymptotic formulae hold,
\bea
\label{smallalpha}
f(\alpha)&\sim&\frac{\pi^{2}}{4\alpha}\qquad\mbox{as $\alpha\ra 0$} \\
\label{largealpha}
f(\alpha)&\sim&\frac{\pi^{2}}{4e_{1}^{2}\alpha^{3}}\qquad\mbox{as $\alpha\ra
\infty$}.
\eea
\end{lemma}
{\bf Proof:} We need only prove (\ref{smallalpha}) since (\ref{largealpha})
follows from this and equation (\ref{fident}). The idea is to split the 
integration region of (\ref{fdef}) into a small neighbourhood of $0$ and its
complement, bound the contribution of the latter region and use a Laurent
expansion in the former.

Fix some $\epsilon\in(0,\frac{1}{4})$ and split $T^{2}$ into $D_{\epsilon}(0)
\coprod(T^{2}\backslash D_{\epsilon}(0))$, where $D_{\epsilon}(0)$ is the open
disk of radius $\epsilon$ centred on $0$. Then
\beq
\int_{D_{\epsilon}(0)}\frac{|\p|^{2}}{(1+\alpha^{2}|\p|^{2})^{2}}<
f(\alpha)<
\int_{D_{\epsilon}(0)}\frac{|\p|^{2}}{(1+\alpha^{2}|\p|^{2})^{2}}+
\int_{T^{2}\backslash D_{\epsilon}(0)}
\frac{|\p|^{2}}{(1+\alpha^{2}|\p|^{2})^{2}}
\eeq
and $|\p|$ is bounded on $T^{2}\backslash D_{\epsilon}(0)$, so there exists
$M_{\epsilon}\in(0,\infty)$ such that
\beq
\int_{T^{2}\backslash D_{\epsilon}(0)}
\frac{|\p|^{2}}{(1+\alpha^{2}|\p|^{2})^{2}}<
\int_{T^{2}\backslash D_{\epsilon}(0)}|\p|^{2}<M_{\epsilon}
\eeq
independent of $\alpha$. Hence
\beq
\int_{D_{\epsilon}(0)}\frac{\alpha|\p|^{2}}{(1+\alpha^{2}|\p|^{2})^{2}}<
\alpha f(\alpha)<\alpha M_{\epsilon}+
\int_{D_{\epsilon}(0)}\frac{\alpha |\p|^{2}}{(1+\alpha^{2}|\p|^{2})^{2}},
\eeq
so it suffices to prove that
\beq
\lim_{\alpha\ra 0}
\int_{D_{\epsilon}(0)}\frac{\alpha |\p|^{2}}{(1+\alpha^{2}|\p|^{2})^{2}}
=\frac{\pi^{2}}{4}.
\eeq
The function $h(z)=z^{2}\p(z)$ is analytic, bounded and (since $s_{0}\notin
D_{\epsilon}(0)$) bounded away from $0$ on $D_{\epsilon}(0)$. So
$\p(z)=h(z)/z^{2}$ where $0<c_{1}<|h(z)|<c_{2}<\infty$, $c_{1}$ and $c_{2}$
being constants. Defining $\gamma=\epsilon/\sqrt{\alpha}$ and 
$u=z/\sqrt{\alpha}$,
\bea
\lim_{\alpha\ra 0}
\int_{D_{\epsilon}(0)}\frac{\alpha |\p|^{2}}{(1+\alpha^{2}|\p|^{2})^{2}}
&=&\lim_{\alpha\ra 0}
\int_{D_{\epsilon}(0)}dz\, d\bar{z}\, \frac{\alpha|h(z)|^{2}/|z|^{4}}{(1+
\alpha^{2}|h(z)|^{2}/|z|^{4})^{2}} \nonumber \\
&=&
\label{l38}
\lim_{\alpha\ra 0}
\int_{\C}du\, d\bar{u}\, \chi_{\gamma}(u)\frac{|h(\sqrt{\alpha}u)|^{2}|u|^{4}}
{(|u|^{4}+|h(\sqrt{\alpha}u)|^{2})^{2}}
\eea
where $\chi_{\gamma}$ is the characteristic function of the disk
(i.e.\ $\chi_{\gamma}(u)=1$ if $|u|<\gamma$, $0$ otherwise). The integrand of
(\ref{l38}) is bounded above, independent of $\alpha$, by
\beq
\frac{c_{2}^{2}|u|^{4}}{(c_{1}^{2}+|u|^{4})^{2}}
\eeq
which is integrable on $\C$. Hence, Lebesgue's dominated convergence theorem
applies \cite{Cho2}, and we may interchange the order of limit and integration
in equation (\ref{l38}). From the Laurent expansion of $\p$ about $0$,
\beq
\p(z)=\frac{1}{z^{2}}+O(z^{2})
\eeq
one sees that $h(z)=1+O(z^{4})$, whence
\beq
\lim_{\alpha\ra 0}
\frac{\chi_{\gamma}(u)|h(\sqrt{\alpha}u)|^{2}|u|^{4}}{(|u|^{4}+
|h(\sqrt{\alpha}u)|^{2})^{2}}=\frac{|u|^{4}}{(1+|u|^{4})^{2}.}
\eeq
Integrating this function over $\C$ yields $\frac{\pi}{4}$, which completes the
proof. $\Box$

There now immediately follows
\begin{theorem} The moduli space $(\M,g)$ is geodesically incomplete.
\end{theorem}
{\bf Proof:} We need only prove that the length of $\Sigma_{V}$,
\beq
l=\int_{0}^{\infty}d\alpha\, \sqrt{f(\alpha)}
\eeq
is finite. By Lemma 2 there exist $0<\alpha_{1}<\alpha_{2}<\infty$,
$0<c_{3},c_{4}<\infty$ such that $f(\alpha)<c_{3}/\alpha$ on $(0,\alpha_{1})$
and $f(\alpha)<c_{4}/\alpha^{3}$ on $(\alpha_{2},\infty)$. Hence,
\bea
l&<&2\sqrt{c_{3}\alpha_{1}}+
\int_{\alpha_{1}}^{\alpha_{2}}d\alpha\, \sqrt{f(\alpha)}+
2\sqrt{\frac{c_{4}}{\alpha}} \nonumber \\
&<& 2\sqrt{c_{3}\alpha_{1}}+(\alpha_{2}-\alpha_{1})f(\alpha_{1})
+2\sqrt{\frac{c_{4}}{\alpha}}
\eea
by monotonicity of $f$. $\Box$

The geodesic approximation predicts, then, that lumps (at least when 
coincident) can shrink and form singularities in finite time. Shrinking has
been observed in numerical simulations of the $\CP$ model in both the plane
\cite{Zak1} and the torus \cite{Zak2}, although the particular initial value
problem considered here has not been simulated.

\section{A two-dimensional geodesic submanifold}
\label{sec:tgs}

The Viergruppe $V_{4}$ has three $\Z_{2}$ subgroups $\{Id,P\}$, $\{Id,R\}$
and $\{Id,PR\}$ whose fixed point sets $\wh{\Sigma}_{P}$, $\wh{\Sigma}_{R}$,
$\wh{\Sigma}_{PR}$ respectively are all geodesic submanifolds of $(\Gh,\gh)$.
Of these, $\wh{\Sigma}_{R}$ is two dimensional (the others are three
dimensional) and projects under $\wt{\phi}$ to
\beq
\Sigma_{R}=\{\alpha e^{i\psi}\p(z):\alpha\in\R^{+},\psi\in[0,2\pi]\},
\eeq
a geodesic submanifold of $(\M,g)$ diffeomorphic to a cylinder. This has
a tractable geodesic problem. Recalling that $g$ is left-invariant under
$SU(2)$, its restriction $g_{R}$ to $\Sigma_{R}$ is independent of $\psi$.
In fact,
\beq
g_{R}=f(\alpha)(d\alpha^{2}+\alpha^{2}d\psi^{2})
\eeq
where $f$ is the same function defined in (\ref{fdef}).

Lemma 2 implies that the asymptotic form of $g_{R}$ towards the ends of the
cylinder is
\bea
\label{a47}
g_{R}&\sim&\frac{\pi^{2}}{4\alpha}(d\alpha^{2}+\alpha^{2}d\psi^{2})\qquad
\mbox{as $\alpha\ra 0$} \\
\label{a48}
g_{R}&\sim&
\frac{\pi^{2}}{4e_{1}^{2}\alpha^{3}}(d\alpha^{2}+\alpha^{2}d\psi^{2})
\qquad\mbox{as $\alpha\ra\infty$}.
\eea
The formula in (\ref{a47}) is the metric of a flat, singular cone with
deficit angle $\pi$, so $(\Sigma_{R},g_{R})$ can be visualized as having
a conical singularity at $\alpha=0$. By virtue of the identity (\ref{fident}),
the metric $g_{R}$ is invariant under the mapping $\alpha\mapsto(e_{1}^{2}
\alpha)^{-1}$, and consequently  $(\Sigma_{R},g_{R})$ has an identical conical 
singularity
at $\alpha=\infty$, as may be shown by the reparametrization $\beta=(e_{1}^{2}
\alpha)^{-1}$ in (\ref{a48}). So $(\Sigma_{R},g_{R})$ is a rotationally
symmetric cylinder of finite length with its ends pinched to identical cones.
It is the internal rotation  orbit, about
a fixed axis, of the one parameter family of exceptionally symmetric 
configurations already described ($\Sigma_{V}$). The singular points $\alpha=
0$ and $\alpha=\infty$ correspond to infinitely narrow, spiky, ring like
configurations centred on $0$ and $s_{0}$ respectively. Motion on 
$\Sigma_{R}$
corresponds to rotational and shape changing motion of the double lump on the 
torus. The conserved kinetic energy of this motion is
\beq
\label{69}
T=f(\alpha)\dot{\alpha}^{2}+\frac{J^{2}}{4\alpha^{2}f(\alpha)},
\eeq
where $J=\alpha^{2}f(\alpha)\dot{\psi}$ is the conserved angular momentum
conjugate to $\psi$. One may imagine the dynamics as that of a point particle
moving on the interval $(0,\infty)$ with position dependent mass and subject
to a potential. Geodesic motion is invariant under rescaling of time, so one 
can restrict attention to the two cases $J^{2}=0$ and $J^{2}=1$. If $J^{2}=0$
the motion is irrotational and the point particle travels from one conical
singularity to the other in finite time along a path of constant $\psi$.
These geodesics are just rotated versions of $\Sigma_{V}$. The more interesting
case is when $J^{2}=1$ where the nature of the motion is determined by the
centrifugal potential
\beq
\label{70}
{\cal U}(\alpha)=\frac{1}{4\alpha^{2}f(\alpha)}.
\eeq
From the asymptotic formulae of proposition 1 we see that the potential has
the asymptotic behaviour
\bea
\label{71}
{\cal U}(\alpha)&\sim& \frac{}{\pi^{2}\alpha}, \quad\mbox{as }\alpha\ra 0 \\
\label{71a}
{\cal U}(\alpha)&\sim&\frac{e_{1}^{2}}{\pi^{2}}\alpha\quad\mbox{as }\alpha
\ra\infty
\eea
implying that ${\cal U}$ must have at least one stable equilibrium. 
The identity (\ref{fident}) implies a similar identity for the potential,
\beq
\label{uident}
{\cal U}(\frac{1}{\alpha e_{1}^{2}})\equiv {\cal U}(\alpha).
\eeq
Differentiating both sides of (\ref{uident}) one finds that ${\cal U}$ has
a critical point at $\alpha=1/e_{1}$, the fixed point of the isometry
$\alpha\mapsto(\alpha e_{1}^{2})^{-1}$. 
Numerical evaluation of ${\cal U}$ suggests that this is the only critical
point, a global minimum, so that ${\cal U}$ is a
single potential well (see figure 6). Since ${\cal U}$ grows unbounded
as $\alpha\ra 0$ and $\alpha\ra\infty$, all motion in the well is
oscillatory. So these geodesics wind around $\Sigma_{R}$, passing back and 
forth along its length indefinitely. They are bounded away from the 
singularities by angular momentum conservation. They correspond to rotational
motions of the double lump during which the arrows of the configuration spin
about the north-south axis of $S^{2}$, and its shape periodically
oscillates about that of the most symmetric configuration, $W(z)=\p(z)/e_{1}$.

\section{The diameter of $(\M,g)$}
\label{sec:dia}

In this section we will prove that $(\M,g)$ has finite size, in an appropriate
sense. Since one is interested in $(\M,g)$ primarily for its geodesics, a
linear measure of size is most meaningful, so we will consider its diameter.
Recall that $(\M,g)$, like any Riemannian manifold, has a natural distance
function $d:\M\times\M\ra\R$ where $d(W,W')$ is the infimum of lengths with 
respect to $g$ of
piecewise $C^{1}$ paths in $\M$ connecting $W$ and $W'$ (note that $d$ has 
nothing to do with $D$, the distance function defined in 
section \ref{sec:mod},
although they define equivalent topologies on $\M$). The diameter of $(\M,g)$
is simply the diameter of the associated metric space $(\M,d)$, that is,
\beq
{\rm diam}(\M,g)=\sup_{W,W'\in\M}d(W,W').
\eeq
Once again, it is the noncompactness of $\M$ which makes this diameter
interesting, and its finiteness nontrivial. The geometric meaning of the
result is that all points lie within a bounded distance of each other, and,
in particular, no point lies far from the end of $\M$, where the static
solutions collapse to singular, spiky configurations. Thus all static
solutions are close to collapse in this geometry. This may be the underlying
cause of the ubiquitous instability found in numerical simulations of two-lump
scattering on the torus \cite{Zak2}.

\begin{theorem} The moduli space $(\M,g)$ has finite diameter.
\end{theorem}
{\bf Proof:} It suffices to prove that the covering space $(G,\g)$ has
finite diameter and, further, since $\g=\gh\oplus\delta$ is a product metric
and $T^{2}$ is compact, it is sufficient to prove that the reduced covering
space $(\Mh,\gh)$ has finite diameter. By the triangle inequality for
$d:\Mh\times\Mh\ra\R,$
\beq
{\rm diam}(\Mh,\gh)\leq 2\sup_{W\in\Mh}d(W,W_{0}),
\eeq
where $W_{0}$ is any point in $\Mh$. Let $W_{0}=\p$. We will explicitly 
construct a path from $W\in\Mh\cong SO(3)\times[\R^{+}\times\C]$ to $W_{0}$,
and bound its length independent of $W$.

Let $W=U\odot[\alpha(\p+\rho)]$. The first piece of the path has $(\alpha,\rho)
\in\R^{+}\times\C$ fixed, but takes $U$ to $I$. For example, since any $U\in
SU(2)$ is $\exp(u)$ for some $u\in su(2)$, we could consider the path
$\Omega(t)=\exp((1-t)u)$, so that $\Omega(0)=U$ while $\Omega(1)=I$. Denote
by $\gamma(\alpha,\rho)$ the metric on $SO(3)$ induced by $\gh$ at fixed
$(\alpha,\rho)$. Since $SO(3)$ is compact the length of $\Omega(t)$ is bounded
independent of $U$ for each $(\alpha,\rho)$. One must check, however, that the
length remains bounded as a function of $(\alpha,\rho)$. Since
$\gamma(\alpha,\rho)$ is a left-invariant metric on $SO(3)$, it suffices to
show that 
\beq
\Gamma(\alpha,\rho):=\sum_{i,j=1}^{3}|\gamma_{ij}(\alpha,\rho)|
\eeq
is a bounded function, where $\gamma_{ij}(\alpha,\rho)$ are the metric
coefficients of $\gamma$ evaluated at $I\in SO(3)$ with respect to a particular
choice of basis for $T_{I}SO(3)$. The basis used does not matter. One
convenient choice consists of the three vectors represented by the curves
\beq
\exp\left(it\frac{\tau_{i}}{2}\right)\qquad i=1,2,3\qquad t\in(-\epsilon,
\epsilon),
\eeq
where $\tau_{i}$ are the Pauli matrices (this is equivalent to choosing
$\{i\tau_{i}/2:i=1,2,3\}$ as a basis for $su(2)$). Elementary calculation then
shows that $\Gamma<3$ for all $(\alpha,\rho)$. For example, $\gamma_{33}
(\alpha,\rho)$ is the squared length of the vector $[\exp(it\tau_{3}/2)]$.
Let $w:=\alpha(\p+\rho)$. Then
\bea
W(z,t)&=&\exp\left(it\frac{\tau_{3}}{2}\right)\odot w(z)=e^{it}w(z) \nonumber
\\
\dot{W}(z)&=&\left.\frac{\cd W}{\cd t}\right|_{t=0}=iw(z)
\eea
and
\beq
\gamma_{33}(\alpha,\rho)=\Tint\frac{|\dot{W}|^{2}}{(1+|W|^{2})^{2}}
=\Tint\frac{|w|^{2}}{(1+|w|^{2})^{2}}<\frac{1}{2}.
\eeq
Bounds on the other metric coefficients are equally straightforward.

It remains to construct a path from $\alpha(\p+\rho)$ to $\p$ with length
bounded above independent of $(\alpha,\rho)$. It is necessary to split
$\R^{+}\times\C$ into two pieces $X_{+}\coprod X_{-}$ and construct the path
differently in each piece. Here $X_{+}=\{(\alpha,\rho)\in\R^{+}\times\C:
\alpha>1\}$ and $X_{-}$ is its complement.

For any $(\alpha,\rho)\in X_{-}$ construct the path $x_{-}:[0,1]\ra\R^{+}\times
\C$ where
\beq
x_{-}(t)=\left\{
\begin{array}{cc}
(\alpha,(1-2t)\rho) & t\in[0,\frac{1}{2}] \\
(1+2(1-\alpha)(t-1),0) & t\in(\frac{1}{2},1] \\
\end{array}\right.
\eeq
so that $x_{-}(0)=(\alpha,\rho)$, $x_{-}(1)=(1,0)$. Thinking of $\R^{+}\times
\C$ as the upper half of $\R^{3}$, this path consists of a horizontal line
from $(\alpha,\rho)$ to $(\alpha,0)$ followed by a vertical line from
$(\alpha,0)$ to $(1,0)$ (see figure 7). Its length is bounded above by
the sum of the lengths of the curves $\{(\alpha,te^{i\psi}):t\in[0,\infty)$
and $\{(t,0):t\in(0,1]\}$, where $\psi={\rm arg}\rho$. So
\beq
l[x_{-}]<l_{1}(\alpha,\psi)+l_{3}
\eeq
where 
\bea
l_{1}(\alpha,\rho)&=&\int_{0}^{\infty}d|\rho|\sqrt{\gh_{|\rho||\rho|}(\alpha,
\rho)}=\int_{0}^{\infty}d|\rho|\left[\Tint\frac{\alpha^{2}}{(1+\alpha^{2}
|\p+|\rho|e^{i\psi}|^{2})^{2}}\right]^{\frac{1}{2}} \nonumber \\
l_{3}&=&\int_{0}^{1}d\alpha\sqrt{\gh_{\alpha\alpha}(\alpha,0)}
=\int_{0}^{1}d\alpha\left[\Tint\frac{|\p|^{2}}{(1+\alpha^{2}|\p|^{2})^{2}}
\right]^{\frac{1}{2}}.
\eea
That $l_{3}$ is finite follows directly from Lemma 2, since $\gh_{\alpha\alpha}
(\alpha,0)$ is precisely $f(\alpha)$, the function previously discussed. To
prove that $l_{1}(\alpha,\rho)$ is finite and bounded independent of
$(\alpha,\psi)\in(0,1]\times[0,2\pi]$ is more involved. 
By a change of variable,
$\sigma:=\alpha|\rho|,$ we can rewrite $l_{1}(\alpha,\psi)$ as
\beq
l_{1}(\alpha,\psi)=\int_{0}^{\infty}d\sigma\left[\Tint\frac{1}{(1+|\alpha\p+
\sigma e^{i\psi}|^{2})^{2}}\right]^{\frac{1}{2}}.
\eeq
One must now appeal to a technical lemma, whose proof we postpone:
\begin{lemma} There exist $\sigma_{*},C>0,$ independent of $(\alpha,\psi),$
such that $\forall\sigma>\sigma_{*}$ and $\alpha\leq 1$,
\beq
\Tint\frac{1}{(1+|\alpha\p+\sigma e^{i\psi}|^{2})^{2}}<\frac{C}{\sigma^{3}}.
\qquad\clubsuit
\eeq
\end{lemma}
It follows that
\bea
l_{1}(\alpha,\psi)&=&\int_{0}^{\sigma_{*}}\left[\Tint\frac{1}{(1+|\alpha\p+
\sigma e^{i\psi}|^{2})^{2}}\right]^{\frac{1}{2}}+
\int_{\sigma_{*}}^{\infty}\left[\Tint\frac{1}{(1+|\alpha\p+
\sigma e^{i\psi}|^{2})^{2}}\right]^{\frac{1}{2}} \nonumber \\
&<&\sigma_{*}+\int_{\sigma_{*}}^{\infty}\sqrt{\frac{C}{\sigma^{3}}}<C'<\infty
\eea
$C'$ being a constant.

Now, for any $(\alpha,\rho)\in X_{+}$ construct the path $x_{+}:[0,1]\ra
\R^{+}\times\C$, where
\beq
x_{+}(t)=\left\{\begin{array}{cc}
(\alpha-2(\alpha-1)t,\rho) & t\in[0,\frac{1}{2}] \\
(1,2(1-t)\rho) & t\in(\frac{1}{2},1], \\
\end{array}\right.
\eeq
consisting (see figure 7) of a vertical line from $(\alpha,\rho)$ to 
$(1,\rho)$ followed by a horizontal line from $(1,\rho)$ to $(1,0)$. Its length
is bounded above by the sum of the lengths of the lines 
$\{(t,\rho):t\in[1,\infty)\}$ and $\{(1,te^{i\psi}):t\in[0,\infty)\}$. So
\beq
l[x_{+}]<l_{2}(\rho)+l_{1}(1,\psi)
\eeq
where $l_{1}$ was previously defined and 
\beq
l_{2}(\rho)=\int_{1}^{\infty}d\alpha\sqrt{\gh_{\alpha\alpha}(\alpha,\rho)}
=\int_{1}^{\infty}d\alpha\left[\Tint\frac{|\p+\rho|^{2}}{(1+\alpha^{2}|\p+
\rho|^{2})^{2}}\right]^{\frac{1}{2}}.
\eeq
We have already shown that $l_{1}(1,\psi)$ is finite and bounded independent
of $\psi$ (this follows from Lemma 3 in the case $\alpha=1$).

That $l_{2}(\rho)$ is finite $\forall\rho\in\C$ is easily shown, using an
argument similar to that of Lemma 2. Let $z_{1}, z_{2}$ be the roots of
$\p+\rho$ (possibly coincident) and split $T^{2}$ into small neighbourhoods of
these roots and their complement. In the complement use the trivial bound
$|\p+\rho|\geq C$, constant, while near the roots use Laurent expansions of
$\p+\rho$. One finds that $\gh_{\alpha\alpha}<C'/\alpha^{3}$ which is
sufficient for finiteness of $l_{2}(\rho)$ for all $\rho$, and boundedness of 
$l_{2}$ on any
compact subset of $\C$. This is insufficient for our purposes, since
$l_{2}(\rho)$ could grow unbounded as $|\rho|\ra\infty$. We again appeal to
a technical lemma whose proof we postpone:
\begin{lemma} For all $\rho\in\C$ such that $|\rho|>e_{1}+2$,
\bea
\Tint\frac{|\p+\rho|^{2}}{(1+\alpha^{2}|\p+\rho|^{2})^{2}}<
\frac{2}{\alpha^{4}}+\frac{\pi}{2\alpha^{4}}\log(1+\alpha^{2}).\qquad\clubsuit
\eea
\end{lemma}
So for all $\rho$ outside the closed disk $D_{e_{1}+2}(0)$,
\beq
l_{2}(\rho)<\int_{1}^{\infty}d\alpha\left[\frac{2}{\alpha^{4}}+\frac{\pi}{2
\alpha^{4}}\log(1+\alpha^{2})\right]^{\frac{1}{2}}=C<\infty
\eeq
$C$ being a constant. Hence $l_{2}$ is bounded independent of $\rho$, and all 
points in $\Mh$ lie within a bounded distance of $\p$, and hence, one another.
$\Box$

\vspace{0.5cm}
\noindent
{\bf Proof of Lemma 4:} Let $\rho\in\C$ such that $|\rho|>e_{1}+2$. Since
$\p$ is an even function,
\beq
\label{**}
\gh_{\alpha\alpha}(\alpha,\rho)=2\int_{H}\frac{|\p+\rho|^{2}}{(1+\alpha^{2}
|\p+\rho|^{2})^{2}}
\eeq
where $H=[0,1)\times[0,\frac{1}{2})$ is the ``half torus'' (the point is that
$\p$ is injective on $H$). Split $H$ into two pieces $H_{+}\coprod H_{-}$
where $H_{+}=\{z\in H:|\p+\rho|>1\}$. Now,
\beq
\label{***}
\int_{H_{+}}\frac{|\p+\rho|^{2}}{(1+\alpha^{2}|\p+\rho|^{2})^{2}}<
\int_{H_{+}}\frac{1}{\alpha^{4}|\p+\rho|^{2}}<\int_{H_{+}}\frac{1}{\alpha^{4}}
<\frac{1}{\alpha^{4}}.
\eeq
To estimate the contribution of the $H_{-}$ region, we perform a variable
change $z\mapsto u=\p(z)$ on $H_{-}$. Since $\p$ is injective on $H_{-}$, this
variable change is well defined provided $\p$ has no critical (i.e.\ double
valency) points in $H_{-}$. The transformed integration range $\p(H_{-})$ is
a closed disk of unit radius centred on $-\rho$, so given that 
$|\rho|>e_{1}+2$, $\p(H_{-})$ contains none of $\{\infty,0,\pm e_{1}\}$, and
hence $H_{-}$ excludes all the double valency points. The Jacobian of the
variable change is $|\p'(z)|^{-2}=|4u(u^{2}-e_{1}^{2})|^{-1}$, so
\beq
\int_{H_{-}}\frac{|\p+\rho|^{2}}{(1+\alpha^{2}|\p+\rho|^{2})^{2}}
=\frac{1}{4}\int_{\p(H_{-})}\frac{du\, d\bar{u}}{|u||u^{2}-e_{1}^{2}|}
\frac{|u+\rho|^{2}}{(1+\alpha^{2}|u+\rho|^{2})^{2}}.
\eeq
Now, for all $u\in\p(H_{-})$, $|u|\geq e_{1}+1>1$, and $|u\pm e_{1}|\geq
||u|-e_{1}|\geq 1$, so
\bea
\int_{H_{-}}\frac{|\p+\rho|^{2}}{(1+\alpha^{2}|\p+\rho|^{2})^{2}}
&<&\frac{1}{4}\int_{\p(H_{-})}du\, d\bar{u}\,
\frac{|u+\rho|^{2}}{(1+\alpha^{2}|u+\rho|^{2})^{2}} \nonumber \\
&=&\frac{\pi}{2}\int_{0}^{1}dx\, \frac{x^{3}}{(1+\alpha^{2}x^{2})^{2}}
\qquad (x:=|u+\rho|) \nonumber \\
&=&\frac{\pi}{4\alpha^{4}}\int_{1}^{1+\alpha^{2}}dy\, \frac{y-1}{y^{2}}
\qquad (y:=1+\alpha^{2}x) \nonumber \\
\label{****}
&<&\frac{\pi}{4\alpha^{4}}\log(1+\alpha^{2})
\eea
Using inequalities (\ref{***}) and (\ref{****}) in equation (\ref{**}), the
result immediately follows. $\Box$

\vspace{0.5cm}
\noindent
{\bf Proof of Lemma 3:} The idea is similar to the proof of Lemma 4:
\beq
\Tint\frac{1}{(1+|\alpha\p+\sigma e^{i\psi}|^{2})^{2}}=
2\left[\int_{H_{+}}\frac{1}{(1+|\alpha\p+\sigma e^{i\psi}|^{2})^{2}}+
\int_{H_{-}}\frac{1}{(1+|\alpha\p+\sigma e^{i\psi}|^{2})^{2}}\right]
\eeq
where again $H_{+}\coprod H_{-}=H$, the half torus, but now
\beq
H_{+}=\{z\in H:|\alpha\p+\sigma e^{i\psi}|>\sigma^{\frac{3}{4}}\}.
\eeq
The $H_{+}$ integral is trivially bounded by $1/\sigma^{3}$. We make the same
variable change $z\mapsto u=\p(z)$ on $H_{-}$. Now $\p(H_{-})$ is a closed
disk of radius $\sigma^{\frac{3}{4}}/\alpha$ centred on 
$\sigma e^{i(\psi+\pi)}/\alpha$. In order that $\p(H_{-})$ contain none of
$\{\infty,0,\pm e_{1}\},$ it suffices that $\sigma\geq\sigma_{c}$ where
$\sigma_{c}$ is the real solution of
\beq
\sigma_{c}-\sigma_{c}^{\frac{3}{4}}=2e_{1}.
\eeq
To see this, note that $\forall u\in\p(H_{-}),$
\beq
|u|\geq\frac{\sigma-\sigma^{\frac{3}{4}}}{\alpha}
\geq \sigma-\sigma^{\frac{3}{4}}  
\geq \sigma_{c}-\sigma_{c}^{\frac{3}{4}} =2e_{1},
\eeq
where the restriction $\alpha\leq 1$ has been used.
So the variable change is well defined provided $\sigma\geq\sigma_{c}$.

Recall that the Jacobian of the transformation is $|4u(u^{2}-e_{1}^{2})|^{-1}$.
Now $\forall u\in\p(H_{-})$, $|u|\geq 2e_{1}$ as shown above. Hence
\beq
|u|^{3}=|u|^{2}|u|\geq 4e_{1}^{2}|u|,
\eeq
and
\bea
|u(u^{2}-e_{1}^{2})|&\geq&||u|^{3}-e_{1}^{2}|u||=|u|^{3}-e_{1}^{2}|u|
\nonumber \\
&\geq&|u|^{3}-\frac{1}{4}|u|^{3}=\frac{3}{4}|u|^{3}.
\eea
Thus,
\beq
\label{*****}
\int_{H_{-}}\frac{1}{(1+|\alpha\p+\sigma e^{i\psi}|^{2})^{2}}
\leq\frac{1}{3}\int_{\p(H_{-})}\frac{du\, d\bar{u}}{|u|^{3}}\,
\frac{1}{(1+|\alpha u+\sigma e^{i\psi}|^{2})^{2}}.
\eeq
Now let $\tilde{\sigma}_{c}$ be the real solution of
\beq
\tilde{\sigma}_{c}-\tilde{\sigma}_{c}^{\frac{3}{4}}=
\frac{\tilde{\sigma}_{c}}{2},
\eeq
and define $\sigma_{*}=\sup\{\sigma_{c},\tilde{\sigma}_{c}\}$. Then,
provided $\sigma\geq\sigma_{*}$, for all $u\in\p(H_{-})$
\beq
|u|\geq\frac{\sigma-\sigma^{\frac{3}{4}}}{\alpha}\geq\frac{\sigma}{2\alpha}.
\eeq
This allows one to estimate the the $|u|^{-3}$ part of the integrand of
inequality (\ref{*****}), which still holds since $\sigma_{*}\geq\sigma_{c}$:
\bea
\int_{H_{-}}\frac{1}{(1+|\alpha\p+\sigma e^{i\psi}|^{2})^{2}}
&\leq& \frac{8\alpha^{3}}{3\sigma^{3}}\int_{\p(H_{-})}
\frac{du\, d\bar{u}}{(1+|\alpha u+\sigma e^{i\psi}|^{2})^{2}} \nonumber \\
&=&\frac{8\alpha^{3}}{3\sigma^{3}}\frac{2\pi}{\alpha^{2}}
\int_{0}^{1}dx\, \frac{x}{(1+x^{2})^{2}}\qquad (x:=|\alpha u+\sigma e^{i\psi}|)
\nonumber \\
&=&\frac{C\alpha}{\sigma^{3}}\leq\frac{C}{\sigma^{3}}.
\eea
The result immediately follows. $\Box$

\section{Conclusion}
\label{sec:conc}

In this paper we have considered the low-energy dynamics of two $\CP$ lumps
moving on a torus in the framework of the geodesic approximation. We have
proved that the degree 2 moduli space $\M$ is homeomorphic to the left coset 
space $G/G_{0}$, where $G$ is the eight-dimensional, noncompact Lie group
$PSL(2,\C)\times T^{2}$ and $G_{0}$ is a discrete subgroup of order 4. This 
result provides a good global parametrization of $\M$ with unconstrained 
parameters, based on the Weierstrass $\p$ function (this situation should be 
compared with other studies where
$\M$ was parametrized using the Weierstrass $\sigma$ function and constrained
parameters \cite{Zak2,Rou}), and allows a systematic description of the
degree 2 static solutions, some of which display four rather than, as one might
expect, two distinct energy peaks. By lifting the metric $g$ on $\M$ defined by
the kinetic energy to $\g$ on the covering space $G$, we showed that the 
dynamics decouples into a trivial ``centre of mass'' motion and a nontrivial
relative motion of the lumps. This reduces the problem to geodesic motion in
a 6-dimensional reduced covering space $(\Gh,\gh)$. Two further 
results were proved concerning the Riemannian geometry of $(\M,g)$, namely that
the moduli space is geodesically incomplete and has finite diameter. These
imply that static lumps can collapse to singularities in finite time, and that 
all static solutions are close to such singularities. In addition, a two
dimensional geodesic submanifold was identified, and its geometry and geodesics
described in detail.

To make further progress in solving the geodesic problem for $(\M,g)$ one would
need to resort to numerical solution of the geodesic equation. Given the
explicit parametrization of $\M$, and that the metric components are integrals
over a compact, two dimensional domain, such numerical work should be
reasonably economical. In particular, there are two 3 dimensional geodesic
submanifolds whose geodesic problems would be well suited to numerical study,
and which should yield interesting lump dynamics. These are $\Sigma_{P}$ and
$\Sigma_{PR}$, the projected fixed point sets of the isometries
$P,PR:\Gh\ra\Gh$. Explicitly,
\bea
\Sigma_{P}&=&\{\exp(i\psi\tau_{2}/2)\odot[\alpha(\p(z)+\rho_{1})]:
\psi\in[0,2\pi],\alpha\in\R^{+},\rho_{1}\in\R\} \nonumber \\
\Sigma_{PR}&=&\{\exp(i\psi\tau_{1}/2)\odot[\alpha(\p(z)+i\rho_{2})]:
\psi\in[0,2\pi],\alpha\in\R^{+},\rho_{2}\in\R\},
\eea
so both are internal rotation orbits, about (different) fixed axes, of the
$\alpha(\p+\rho)$ family, but with $\rho$ real ($\Sigma_{P}$) or purely
imaginary ($\Sigma_{PR}$). On these submanifolds, therefore, the two lumps,
when distinct, are constrained to lie either on the central cross and
boundary of the unit square, or its diagonals, respectively. In either case,
they can only scatter through $90^{\circ}$. In the case of $\Sigma_{P}$, for
example, any geodesic which punctures any of the cylinders $\rho_{1}=0$,
$\rho_{1}=e_{1}$, $\rho_{1}=-e_{1}$ at $\alpha$ much greater than 
$1/e_{1}$ gives rise to
$90^{\circ}$ scattering of the lumps. Similarly, any geodesic which punctures
the $\rho_{2}=0$ cylinder in $\Sigma_{PR}$ gives rise to $90^{\circ}$ 
scattering
along the diagonals. Both these processes have been observed in numerical
simulations of the field equation \cite{Zak2}. To understand the long time
behaviour of the geodesics after the scattering event would require
detailed numerical work. 

Other extensions of the present work would be interesting. One can extend the 
geodesic incompleteness results proved here for $(\M,g)$ on $T^{2}$ and 
elsewhere \cite{Spe} for $(M_{1},g)$ on $S^{2}$ to the general
setting of $(M_{n},g)$ for the $\CP$ model on an arbitrary compact Riemann
surface \cite{Sad}. It may well be possible to similarly extend our result
concerning the finite diameter of moduli space to the general setting. 
Also, one would expect that $(\M,g)$ has finite volume, as well as diameter
(although neither guarantees the other), and perhaps this can be established by
making refined versions of estimates such as those in Lemmata 2, 3 and 4.
Finally, it should be emphasised that all our results concern an {\em 
approximation} to the field theory. While this has proved remarkably successful
in all situations where it has been tested, one would ideally like rigorous
analysis to back up physical intuition. Given the singularity of the geometry
of moduli space for the planar $\CP$ model, the model on the torus provides
an ideal starting point for an analysis fashioned after Stuart's work on
vortices and monopoles \cite{Stu}.

\vspace{0.5cm}
\noindent
{\bf Acknowledgments:} I would like to thank Sharad Agnihotri, Jay Handfield
and Carlo Morpurgo for several helpful discussions.

\newpage
\noindent
\large
{\bf Figure captions}
\normalsize
\vspace{1cm}
\newline Figure 1: The fundamental domain of the Weierstrass $\p$ function:
$\p$ is real on the solid lines and imaginary on the dashed lines. The four
double valency points are marked by circles.
\vspace{0.5cm}
\newline Figure 2: Energy density plots of $W(z)=\p(z)+\rho$ for various
values of $\rho$. In plot (a) $\rho=1-i$, so the roots of $W$ are separate
and two lumps form. In plots (b), (c) and (d), $\rho=0,e_{1},-e_{1}$ 
respectively so the
roots of $W$ coincide. Here the energy distribution is roughly annular, 
centred on the double valency points $s_{0},s_{2},s_{1}$.
\vspace{0.5cm}
\newline Figure 3: Energy density plots of $W(z)=\alpha(\p(z)+1-i)$ in the
cases of (a) large $\alpha$ ($\alpha=2$) and (b) small $\alpha$ 
($\alpha=0.03$).
\vspace{0.5cm}
\newline Figure 4: The exceptionally symmetric family $W(z)=\alpha\p(z)$.
The parameter values are (a) $\alpha=4$, (b) $\alpha=0.3$, 
(c) $\alpha=1/e_{1}$, (d) $\alpha=0.01$ and  (e) $\alpha=0.005$. Plot (c)
depicts the most evenly spread energy distribution possible for a degree 2
static solution.
\vspace{0.5cm}
\newline Figure 5: Energy density plot of $W(z)=(\p(z)-i)/e_{1}$.
\vspace{0.5cm}
\newline Figure 6: The centrifugal potential ${\cal U}(\alpha)$
of equation (\ref{70}), solid line, compared with the asymptotic formulae
for ${\cal U}$ for small and large $\alpha$ given in equations
(\ref{71}) and (\ref{71a}), dashed lines.
\vspace{0.5cm}
\newline Figure 7: The paths $x_{-}$ and $x_{+}$ constructed in the proof
of Theorem 3.

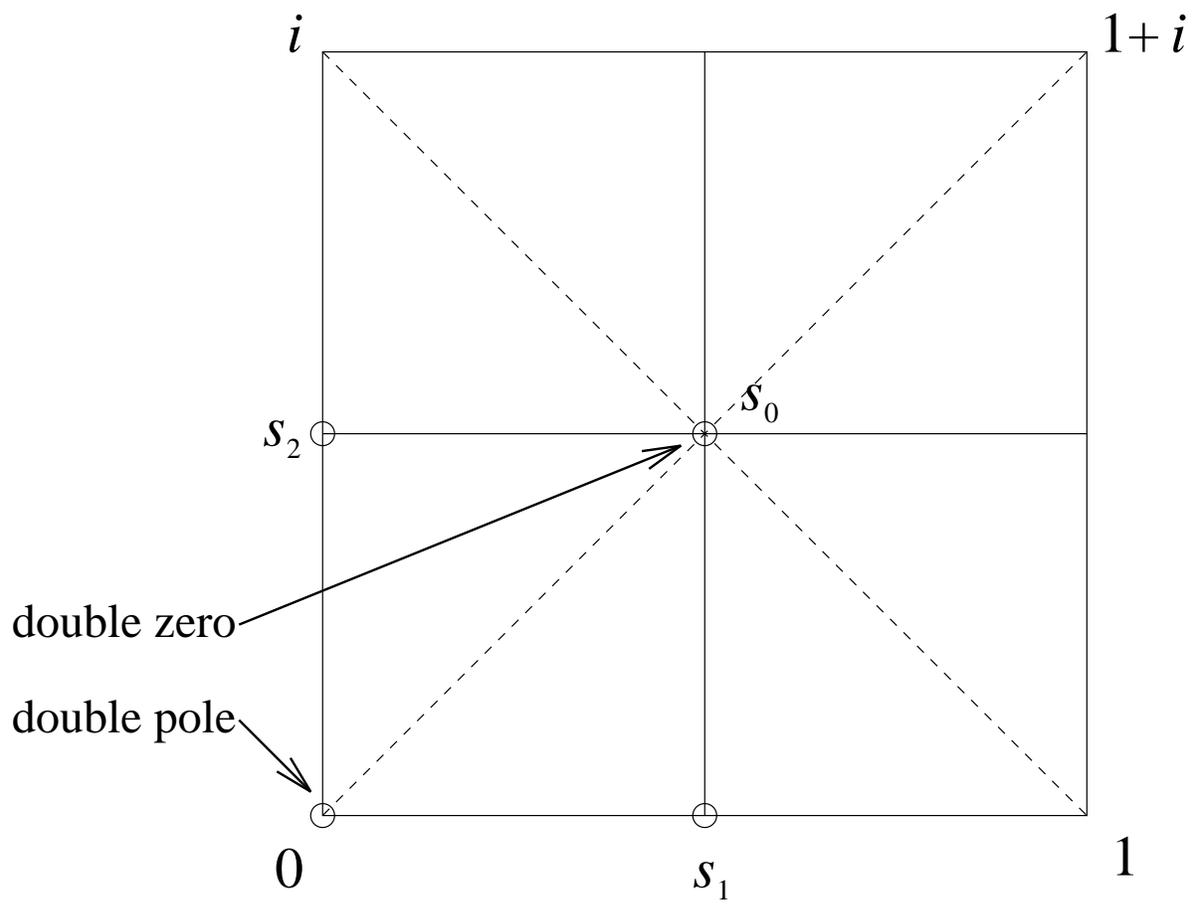
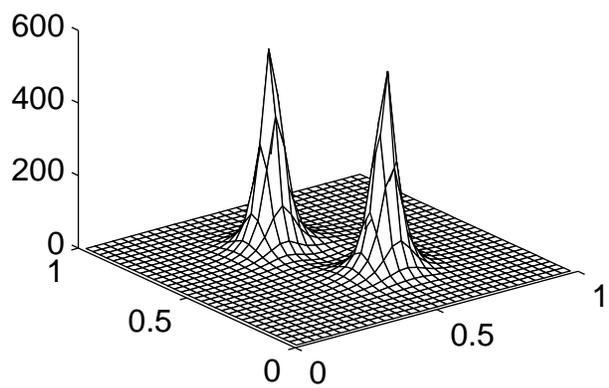
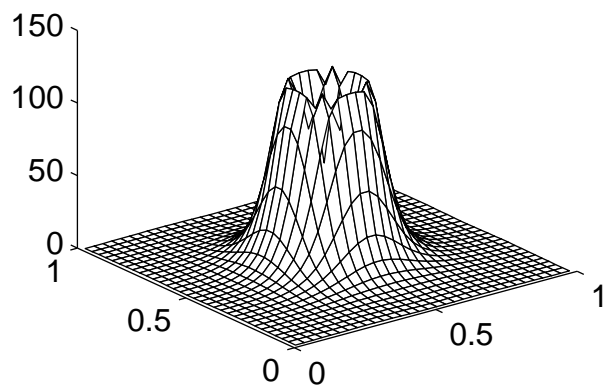
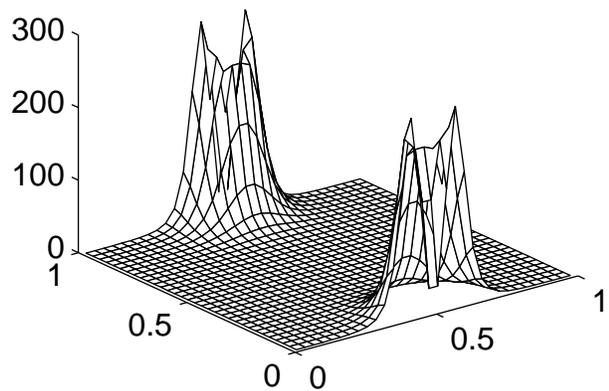
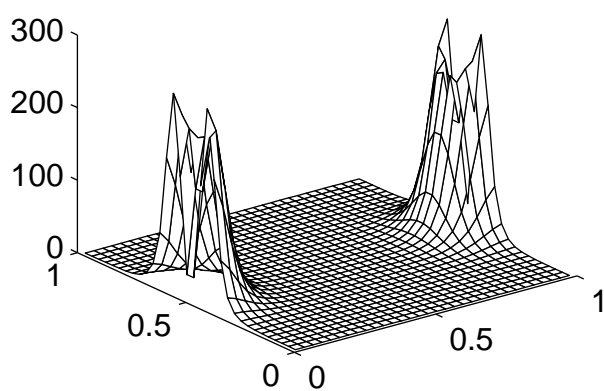
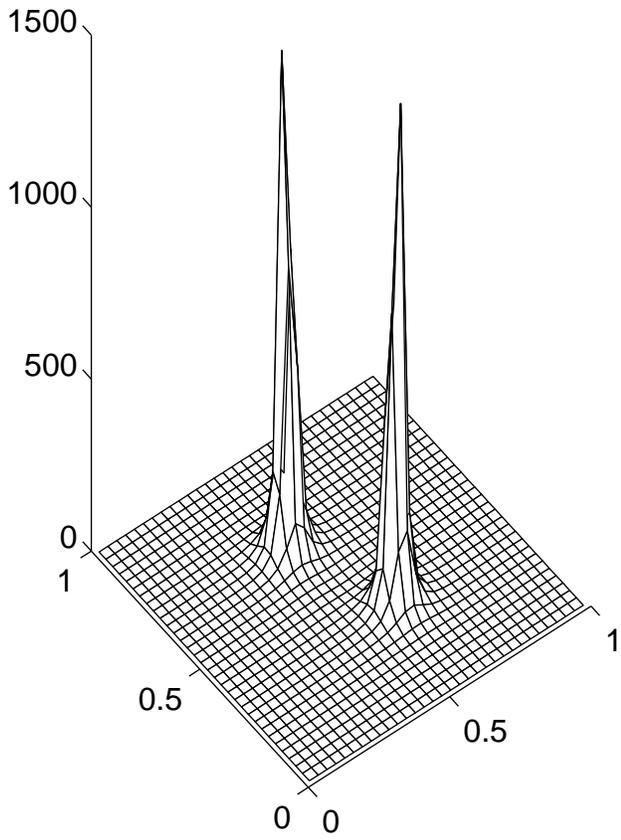
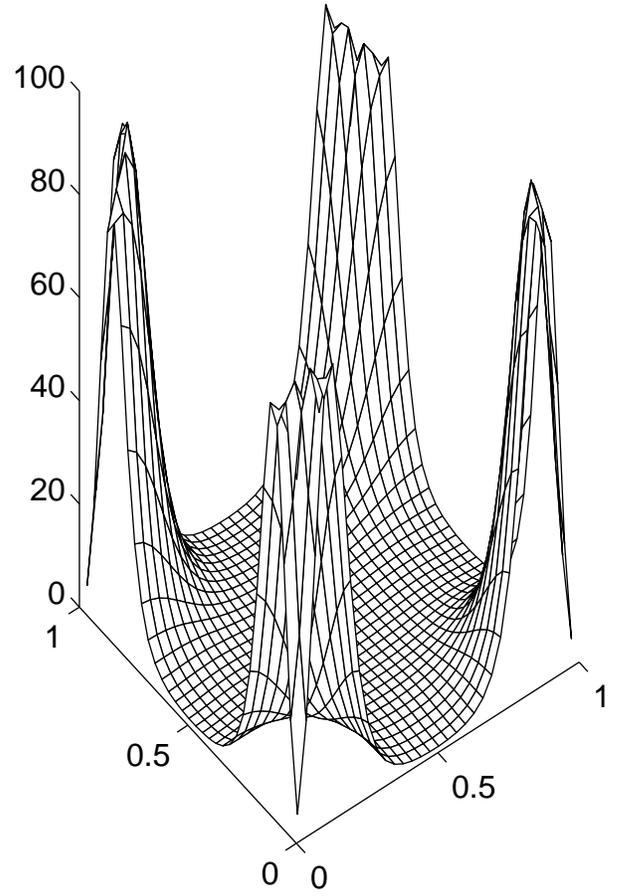
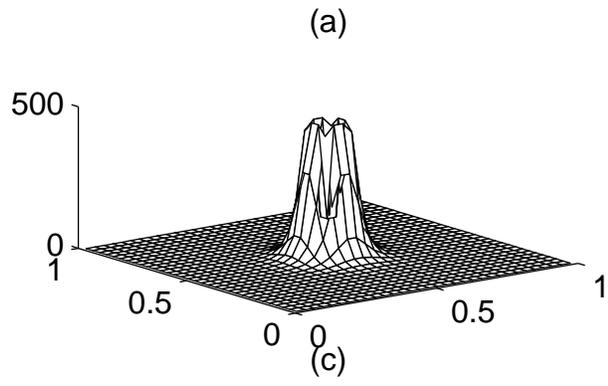
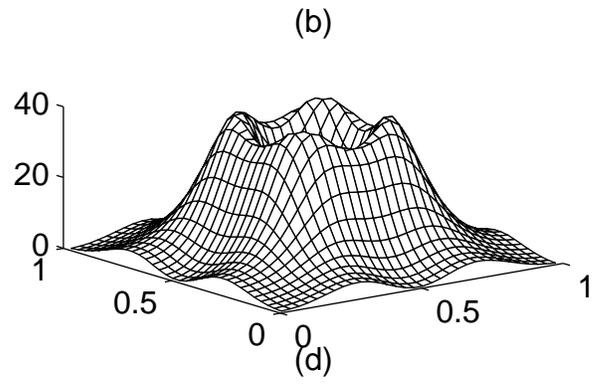
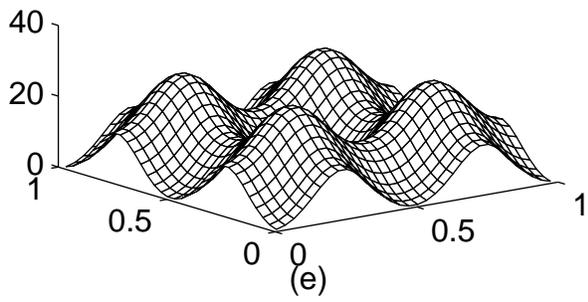
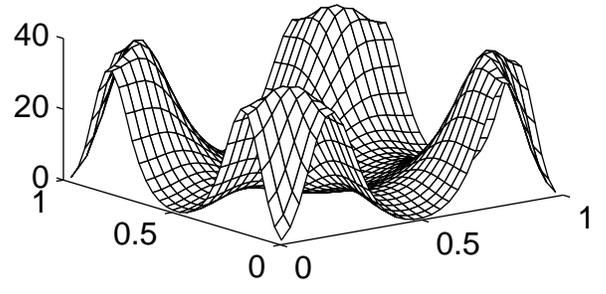
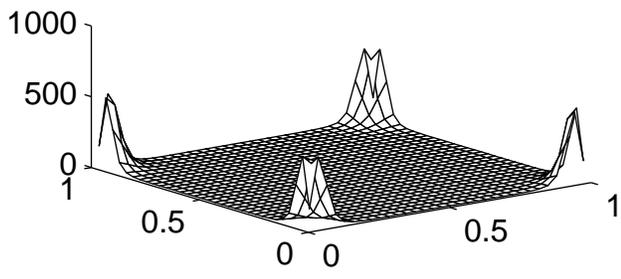
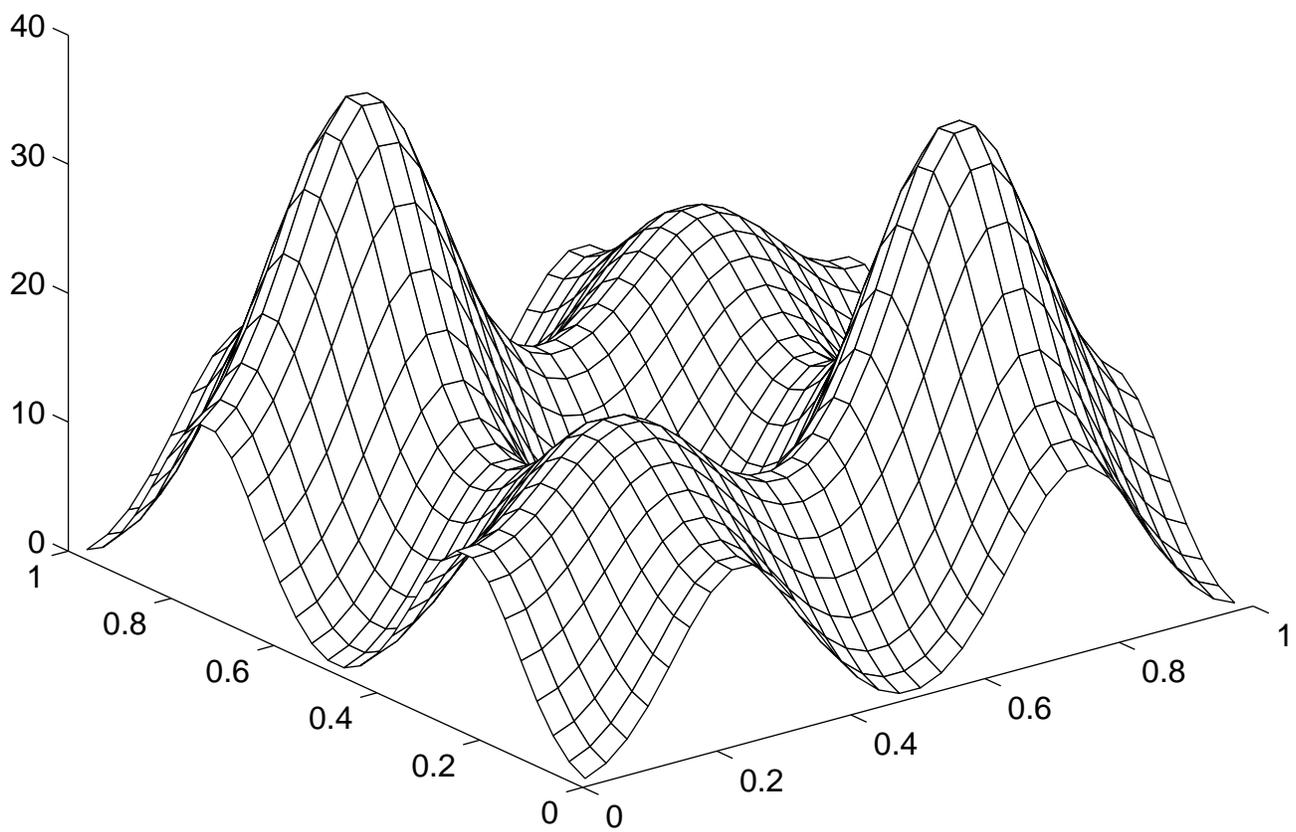
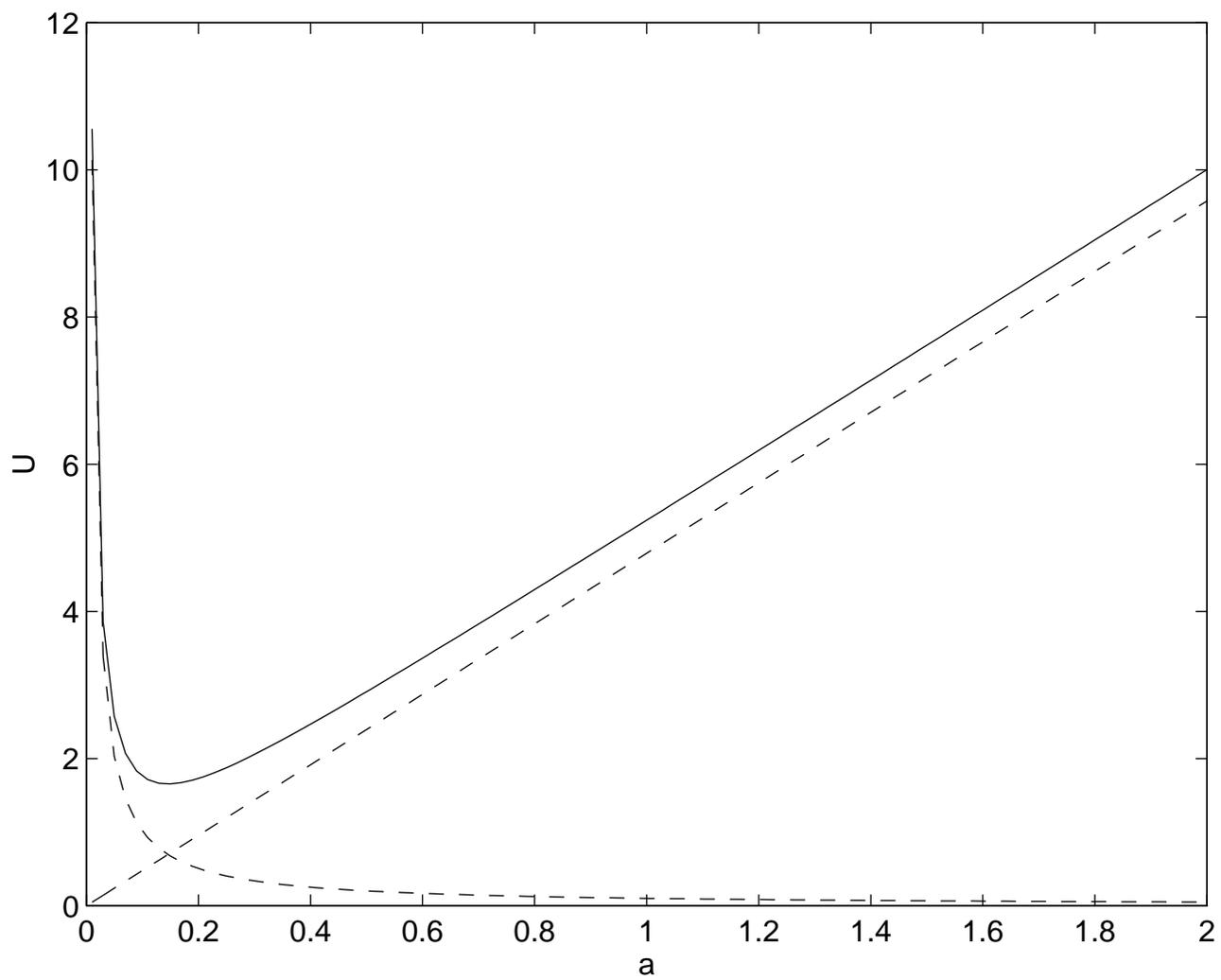
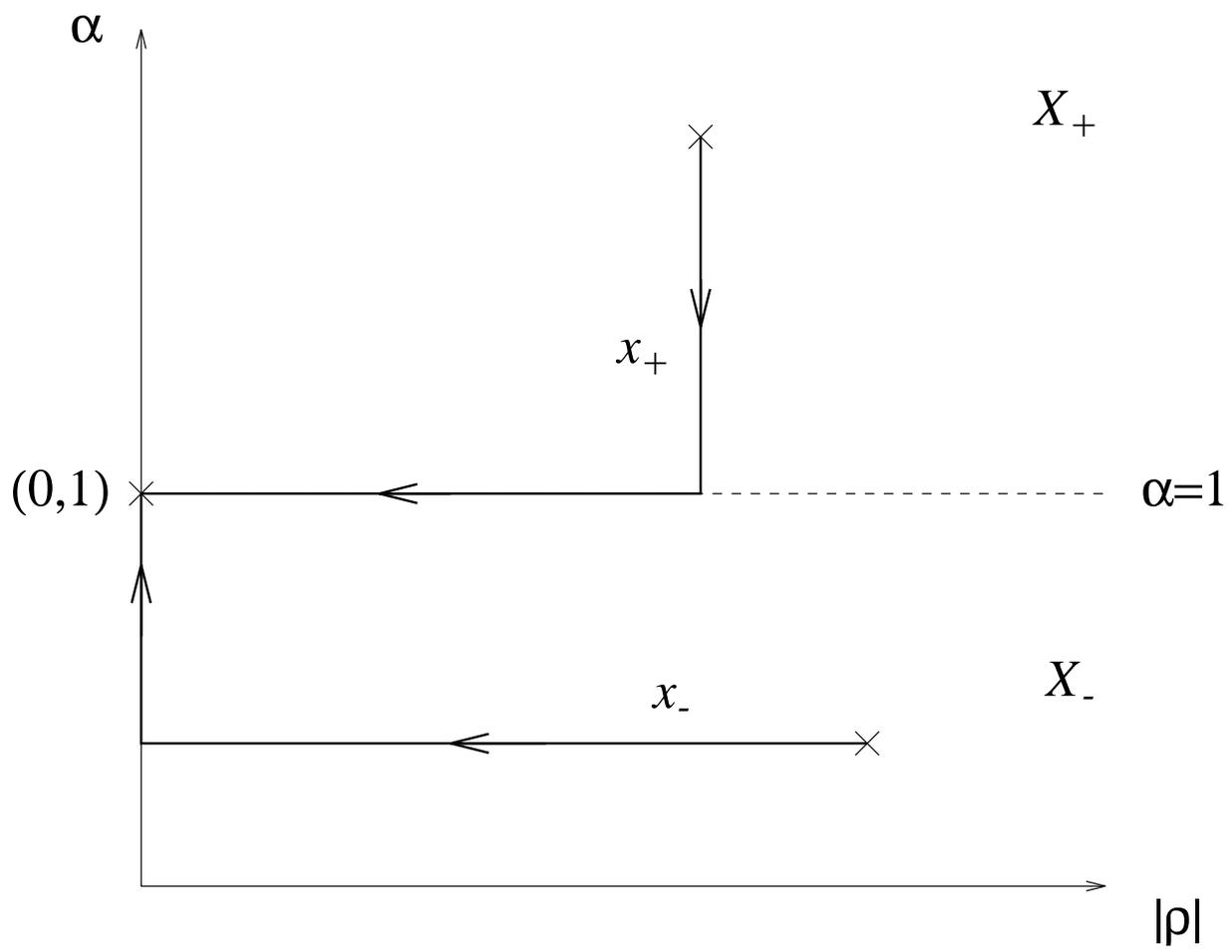
\end{document}